\documentclass[prd,aps,showpacs,preprintnumbers,nofootinbib]
{revtex4}
\usepackage{graphics,graphicx}
\usepackage{amsmath}
\usepackage[T1]{fontenc}
\usepackage{psfrag}
\usepackage{dcolumn}
\usepackage[usenames]{color}
\usepackage{fancybox}
\usepackage{amsfonts}
\allowdisplaybreaks

\definecolor{RedViolet}{rgb}{1.0, 0.0, 0.5}      

\newcommand{\mtot}{m_t}
\newcommand{\OTS}{\sqrt{1-e_t^2}}
\newcommand{\En}{|E|}
\newcommand{\mEn}{{(-E)}}
\newcommand{\peri}{{\bf k}}
\newcommand{\Mean}{{\bf M}}
\newcommand{\MeMo}{{n}}
\newcommand{\magnetic}{m}

\newcommand{\order}[2]{{\cal O}({#1}^{#2})}
\newcommand{\Max}{{\cal M}}

\newcommand{\SA}[2]{{S}^{(#1)}_{#2}}
\newcommand{\CKE}[2]{{\cal \bar \gamma}^{#1}_{#2}}
\newcommand{\SKE}[2]{{\cal \bar \sigma}^{#1}_{#2}}

\newcommand{\vKE}[1]{G_{#1}}

\newcommand{\SAKE}[2]{{\cal S}^{(#1)}_{#2}} 
\newcommand{\AKE}[2]{{\cal A}^{(#1)}_{#2}} 

\newcommand{\ProdSS}[1]{P^{SS, [n]}_{#1}}
\newcommand{\ProdCS}[1]{P^{CS, [n]}_{#1}}
\newcommand{\funcAnq}[2]{{\cal  F}_{[#1 #2]}(u)}
\newcommand{\funcSinAnq}[2]{{\cal F}_{S[#1 #2]}(u)}

\newcommand{\funcSinAKEc}[1]{	({\cal \bar S}_c)^{[#1]}} 
\newcommand{\funcSinAKEs}[1]{	({\cal \bar S}_s)^{[#1]}} 

\newcommand{\funcAKEc}[1]{	({\cal \bar A}_c)^{[#1]}} 
\newcommand{\funcAKEs}[1]{	({\cal \bar A}_s)^{[#1]}} 

\newcommand{\salpha}{{\tilde \alpha}}
\newcommand{\sbeta}{{\tilde \beta}}
\newcommand{\sgamma}{{\tilde \gamma}}
\newcommand{\sdelta}{{\tilde \delta}}
\newcommand{\szeta}{{\tilde \zeta}}

\newcommand{\FourierSin}[7]{{\sum_{j=1}^{\infty} \sin{j\Mean} \Biggl\{ \sum_{k=#2}^{#3} #6_{k#1} \funcAKEs{k,#1}   +   \sum_{k=#4}^{#5} #7_{k#1} \funcSinAKEs{k#1}\Biggr\} }}
\newcommand{\FourierCos}[7]{{\sum_{j=0}^{\infty} \cos{j\Mean} \Biggl\{ \sum_{k=#2}^{#3} #6_{k#1} \funcAKEc{k,#1}   +   \sum_{k=#4}^{#5} #7_{k#1} \funcSinAKEc{k#1}\Biggr\} }}

\begin{document}
\title{Full-analytic frequency-domain 1pN-accurate gravitational wave forms from eccentric compact binaries}
\author{Manuel Tessmer and Gerhard Sch\"afer}
\email{M.Tessmer@uni-jena.de}
\affiliation{Theoretisch-Physikalisches Institut,
Friedrich-Schiller-Universit\"at Jena,
Max-Wien-Platz 1,
07743 Jena, Germany}
\date{\today}

\date{\today}

\begin{abstract}
The article provides ready-to-use 1pN-accurate frequency-domain
gravitational wave forms for eccentric nonspinning compact binaries of
arbitrary mass ratio including the first post-Newtonian (1pN) point particle corrections
to the far-zone gravitational wave amplitude, given in terms of
tensor spherical harmonics. The averaged equations for the decay
of the eccentricity and growth of radial frequency due to radiation reaction
are used to provide stationary phase approximations to the frequency-domain wave forms.
\end{abstract}


\maketitle

The post-Newtonian (pN) description of the dynamics of  compact
binary systems is the topic of actual research. Due to the
strong nonspherically symmetric gravitational interactions, those
objects are supposed to be ``secure'' sources for the detection of
gravitational waves.
Currently, LIGO, VIRGO, and GEO600 search for the last
seconds or minutes in the life of those sources when the
gravitational wave (GW) emission
frequency enters the bandwith of the mentioned detectors. 

The computer resources -- in contrast -- are currently
unable to create numerical GW templates for the early stage of the
binary inspiral (this case is important for LISA), where
hundreds or even thousands of GW cycles have to be simulated
which makes it necessary to propose an analytical prescription
of the orbital evolution, and in the
case of spinning compact binaries, the spin evolution as well.
An equally essential ingredient of the GW data analysis is the
transformation of the GW signal into the frequency domain. There
exist numerous more or less optimized numerical routines to
convert the time domain signal into the frequency or Fourier
domain. To economize on computer resources, it is also
reasonable and desirable to provide analytical Fourier domain
wave forms for the data analysis community. 
Numerous authors have investigated the performance of
circular inspiral templates and their analytical Fourier-domain
pendant, which have been set up to a certain standard in the
literature thanks to the work of Damour, Blanchet and Iyer
to compute higher pN order corrections to
the GW energy loss
\cite{Blanchet:Damour:1986,
Blanchet:Damour:Iyer:1995,
Blanchet:Faye:Iyer:Joguet:2002,
Blanchet:Faye:Iyer:Joguet:2002:err}.

This work is inspired by \cite{Galtsov:Matiukhin:Petukhov:1980} and \cite{Yunes:Arun:Berti:Will:2009}
and is a direct sequel of  \cite{Tessmer:Gopakumar:2006}.
Galt'sov et al. provided a formal frequency-domain decomposition
in multiples of two irreducible frequencies:
the {\em radial} frequency $f_r$, describing the elapsed time
form one periastron passage to the next, and the frequency
associated to the periastron advance parameter, $f_\phi$, but
used this decomposition only for the computation of the far-zone
energy flux, where some mathematical aspects made it unnecessary
to compute the Fourier coefficients directly.

Ref. \cite{Tessmer:Gopakumar:2006} provided numerical insights
to the case of no radiation reaction (RR) with the use of essentially
the same frequencies that appeared in \cite{Galtsov:Matiukhin:Petukhov:1980}.

Yunes et al. \cite{Yunes:Arun:Berti:Will:2009} established analytic  Fourier-domain inspiral
templates for eccentric binary emission for the exemplary
case of Newtonian equations of motion (EOM) and the leading-order GW  amplitude.
Those authors used the expansion of the Kepler equation
and related trigonometric function combinations of the
eccentric anomaly in terms of Bessel functions and applied
this to a series expansion of the leading-order GW amplitude
in the orbital eccentricity.

What we like to do in this article is the generalisation
of Yunes et al.  to 1pN order of the conservative orbital dynamics.
The work will include, in addition to Yunes et al., the
effect of periastron advance and also the GW amplitude 
corrections to 1pN order relative to the leading order
quadrupole approximation.
We present the results in terms of irreducible positive
frequencies $f_r$ and $f_\phi$.
Afterwards, we compute the Fourier domain for the case
when RR is taken into account.

The paper is organized in the following way.
Section \ref{Sec:orbit} gives an overview over the solution
to the 1pN-accurate orbital motion.
Section \ref{Sec::Multipoles} summarizes the transverse-traceless (TT)
projection of the far-zone GW field up to 1pN order corrections
to the leading-order quadrupole field.
The vital components of the Fourier
decomposition  of the GW field are summarised in
Section \ref{Sec::Ingredients}.
Section \ref{Sec::decomp_elements} depicts how to
decompose the terms of the multipole moments of equal
structure.
The Fourier domain multipoles themselves, incorporating
purely 1pN conservative dynamics,
are given in \ref{Sec::Fou_Decomp_Multipoles}.
%
%
The effect of RR is incorporated in Section \ref{Sec::SPA},
using the method of the stationary phase approximation (SPA).

\section{Orbital Motion of compact binaries}
\label{Sec:orbit}
Before we start computing the GW forms, we
give an overview on the solution to the orbital EOM
in a quasi-Keplerian parameterization (QKP).  The QKP
for non-spinning compact binaries up to and including
3pN point particle (PP) contributions can be found in
\cite{Memmesheimer:Gopakumar:Schafer:2004} and
references therein.
We restrict ourselves to the 1pN part thereof.
The QKP for nonspinning compact binaries to 1pN is the
following:
 \begin{eqnarray}
	r &=& a_r (1- e_r \cos u) \,, \label{eq::radialparm}	\\
\Mean :=	n \left(t-t_0\right)  &=& u - e_t \sin u \,, 
\label {eq::KE}\\
	\phi-\phi_0 &=& (1+\peri)\,v 	\,,	\label{eq::angleparm} \\
 v &=& 2\, \arctan \left [ {\sqrt{\frac{1+e_{\phi}}{1-e_{\phi}}} \, \tan \frac{u}{2}} \right ] \label{eq::vparm} \,,	\\
n & = & 2 \sqrt{2} |E|^{3/2}+ \epsilon^2 \frac{|E|^{5/2} (\eta -15) }{\sqrt{2}} \label{eq::n_of_E} \,, \\
 a_r		&=& \frac{1}{2|E|} \, \left\{1	+ \epsilon^2 |E| \left( \frac{\eta-7}{4}	\right)	\right\}		\,,	\\
 e_r		&=& { e_t\, \left\{ 1	+ \epsilon^2 |E| \left( 8-3 \eta			\right) 	\right\} 	}	\,,	\\
 e_\phi	&=& { e_t\, \left\{ 1	+ \epsilon^2 |E| \left( -2 (\eta -4)		\right)	\right\}	}	\,,	\\
 \peri	&=& \epsilon^2 3\,\frac{ \MeMo^{2/3}}{1-e_t^2}=
	   \epsilon^2 \, \frac{6 |E|}{1-e_t^2}\,,	\label{Eq::k_def}		
\end{eqnarray}
In the above Eqs.~(\ref{eq::radialparm})-(\ref{Eq::k_def}),
$\epsilon$ counts the inverse order of the speed of light $c$, $\epsilon^2=c^{-2}$,
$r$ is the scaled relative separation of the two bound objects,
$\phi$ is the orbital phase,
$v$ is the true anomaly,
$\peri$ is the periastron advance parameter,
$e_r$ and $e_t$ are some ``radial'' and ``time''
eccentricity. The latter appears in the Kepler
equation (KE), Eq.~(\ref{eq::KE}), connecting the mean anomaly $\Mean$
(which is directly proportional to the elapsed time $t$ with the proportionality factor $n=2 \pi / P_r= 2 \pi f_r$,
where $P_r$ is the time between two {consecutive} periastron passages)
to the eccentric anomaly $u$,  see \cite{Memmesheimer:Gopakumar:Schafer:2004} for
geometrical insight.
The shorthands $\eta$, $\mu$, and $|E|$ denote the symmetric
mass ration $\eta=(m_1 m_2)/(m_1+m_2)^2$
($m_1$ and $m_2$ are the individual masses with $\mtot=m_1+m_2$), the reduced mass, and the
absolute value of the orbital binding energy.

This  parameterization is essential for the computation
of  the  higher-order time derivatives of the radiative
multipole moments.
We can easily obtain them using the
KE.
 How the orbital parameterization effects the far-zone
gravitational field will be depicted in the next sections.
\section{The 1pN-accurate transverse and traceless far-zone radiation field for eccentric compact binaries}
\label{Sec::Multipoles}
We start collecting the expressions of the transverse-traceless far-zone  field $h^{\rm TT}$ in
Junker and Sch\"afer \cite{Junker:Schafer:1992}. 
They are given in tensor-spherical harmonics and it is remarkable to
note their structure as they appear as
$I^{(l)}_	{  l\magnetic}			=e^{-i \magnetic \phi} \, f_{l\magnetic}(u(t))$ and
$S^{(l)}_	{ l\magnetic}		=e^{-i \magnetic \phi} \, g_{l\magnetic}(u(t))$.
Taking as basis
\begin{eqnarray}
 h^{\rm TT}_{ij}
&=&
\epsilon^4
\frac{G}{R} 
\Biggl\{
						\sum_{\magnetic=-2}^{2} {\stackrel{(2)}{I}}_{2\magnetic} \, T^{E2,2\magnetic}_{ij}
+	\epsilon		\Biggl[	\sum_{\magnetic=-2}^{2} {\stackrel{(2)}{S}}_{2\magnetic} \, T^{B2,2\magnetic}_{ij}
					+	\sum_{\magnetic=-3}^{3} {\stackrel{(3)}{I}}_{3\magnetic} \, T^{E2,3\magnetic}_{ij}	 \Biggr]
\nonumber \\
&&
+	\epsilon^2	\Biggl[	\sum_{\magnetic=-3}^{3} {\stackrel{(3)}{S}}_{3\magnetic} \, T^{B2,3\magnetic}_{ij}
					+	\sum_{\magnetic=-4}^{4} {\stackrel{(4)}{I}}_{4\magnetic} \, T^{E2,4\magnetic}_{ij}
			\Biggr]
\Biggr\}
\,,
\label{Eq::htt_multipoles_1}
\end{eqnarray}
where $R$ is the distance from the observer to the source, $T^{E2,l\magnetic}_{jk}$ and $T^{B2,l\magnetic}_{jk}$ are the ``electric'' and ``magnetic''
tensor spherical harmonics \cite{Thorne:1980}, the superscript $(i)$ allowing $i=2,3,4$ denotes the $i^{\rm th}$ time differentiation.
They read  -- up to the required overall order of $\epsilon^6 = c^{-6}$ in Eq. (\ref{Eq::htt_multipoles_1}) --
 \begin{subequations}
 \begin{align}
  \stackrel{(2)}{I}_{22} =&
8 \sqrt{\frac{2 \pi }{5}} {E} \mu  e^{-2 i \phi }
\Biggl\{
\frac{2-2 e_t^2}{A^2}-\frac{1}{A}+1
	+{\En} \epsilon ^2 \Biggl[\frac{2 (15 \eta
		-82) \left(e_t^2-1\right)}{21 A^3}+\frac{\frac{3}{7} (3 \eta -1) e_t^2+7 \eta -15}{A^2}
\nonumber \\ & 
		+\frac{3 (3 \eta -1)}{14 A}+\frac{3}{14}(1-3 \eta )\Biggr]
\Biggr\}
\nonumber \\ &
{
+i\,
\Biggl\{
\frac{2 {\OTS} e_t \sin (u)}{A^2}+{\En} \epsilon ^2
   \left[\frac{2 {\OTS} (79-6 \eta ) e_t \sin (u)}{21
   A^3}+\frac{e_t \sin (u) \left((9 \eta -3) e_t^2+19 \eta
   -25\right)}{7 A^2 {\OTS}}\right]
\Biggr\}
}
\,,
\label{Def::I222}
\\
\hline
\stackrel{(2)}{I}_{21} =&0
\,,\\
\hline
\stackrel{(2)}{I}_{20} =&
-16 \sqrt{\frac{\pi }{15}} {E} \mu 
\Biggl\{
1-\frac{1}{A(u)}
+{\En} \epsilon ^2 \left[\frac{2 (\eta -26) \left(e_t^2-1\right)}{7 A(u)^3}+\frac{4-19 \eta }{7
   A(u)^2}+\frac{51 \eta -115}{14 A(u)}-\frac{3}{14} (3 \eta -1)\right]
\Biggr\}
\,, 
\label{Def::I202}
\\
\hline
\stackrel{(2)}{S}_{22}=&0
\,,\\
\hline
\stackrel{(2)}{S}_{21} =&\frac{32 \sqrt{\frac{\pi }{5}} \left(m_1-m_2\right) {\mEn}^{3/2}  \eta  e^{-{i\phi}} \sqrt{1-e_t^2}}{3 A(u)^2}\,, \\
\hline
\stackrel{(2)}{S}_{20}=&0
\,, \\
\hline
\stackrel{(3)}{I}_{33}=&
8 \sqrt{\frac{2 \pi }{21}} (-{E})^{3/2} \left(m_1-m_2\right) \epsilon  \eta  e^{-3 i \phi }
\Biggl\{
e_t \sin (u) \left(\frac{4 \left(e_t^2-1\right)}{A(u)^3}-\frac{1}{A(u)}\right)
\nonumber \\
&
+i\,\sqrt{1-e_t^2} \left(\frac{4-4 e_t^2}{A(u)^3}+\frac{3}{A(u)}-\frac{5}{2 A(u)^2}\right)
\Biggr\}
\,, \\
\hline
\stackrel{(3)}{I}_{32}=&0
\,, \\
\hline
\stackrel{(3)}{I}_{31}	=&8 \sqrt{\frac{2 \pi }{35}} (-{E})^{3/2} \left(m_1-m_2\right) \eta  e^{-i \phi } \left\{\frac{{e_t} \sin (u)}{A(u)}-\frac{i \sqrt{1-{e_t}^2}
   \left(1-\frac{5}{6 A(u)}\right)}{A(u)}\right\}
\,, \\
\hline
\stackrel{(3)}{I}_{30}	=& 0\,, \\
\hline
\stackrel{(3)}{S}_{33}	=& 0\,, \\
\hline
\stackrel{(3)}{S}_{32}	=&
e^{-i \,2\,\phi}\,
\frac{8}{3} \sqrt{\frac{2 \pi }{7}} E^2 \sqrt{1-e_t^2} \mu  (1-3 \eta )
 \frac{\left({e_t} \sin (u)-4 i \sqrt{1-{e_t}^2}\right)}{A(u)^3}
\,, \\
\hline
\stackrel{(3)}{S}_{31}	=&	0	\,,	\\
\hline
\stackrel{(3)}{S}_{30}	
=& - 16 \sqrt{\frac{\pi }{105}} {E}^2 {e_t} \sqrt{1-{e_t}^2} \mu  (1-3 \eta ) \frac{ \sin (u)}{A(u)^3}\\
\hline
\stackrel{(4)}{I}_{44}	=&
\frac{4}{9} \sqrt{\frac{2 \pi }{7}} {E}^2 
e^{-4 i \phi }
 \mu (1-3 \eta )
   \Biggl\{\frac{48 \left(1-{e_t}^2\right)^2}{A(u)^4}-\frac{27
   \left(1-{e_t}^2\right)}{A(u)^3}+\frac{43-48 {e_t}^2}{A(u)^2} -\frac{6}{A(u)}+6
\nonumber \\ &
  +\frac{6 i {e_t} \sqrt{1-{e_t}^2} \sin (u) \left(\frac{8
   \left(1-{e_t}^2\right)}{A(u)^2}+\frac{1}{A(u)}+4\right)}{A(u)^2} \Biggr\}
\,,
\\
\hline
\stackrel{(4)}{I}_{43}	=&	0 \,,		\\
\hline
\stackrel{(4)}{I}_{42}	=&	
\frac{8}{63} \sqrt{2 \pi } E^2 \mu  (1-3 \eta )
e^{-2 i \phi }
\Biggl\{-\frac{7-12 {e_t}^2}{A(u)^2}+\frac{3 \left(1-{e_t}^2\right)}{A(u)^3}+\frac{6}{A(u)}-6
-\frac{3 i {e_t}  \sqrt{1-{e_t}^2} \left(\frac{1}{A(u)}+4\right) \sin (u) } {A(u)^2}
\Biggr\}
\,,
\\
\hline
\stackrel{(4)}{I}_{41}	=&	0	\,,	\\
\hline
\stackrel{(4)}{I}_{40}	=&
\frac{8}{21} \sqrt{\frac{\pi }{5}} {E}^2 \mu  (1-3 \eta ) \left\{ \frac{5 \left(1-{e_t}^2\right)}{A(u)^3}-\frac{6}{A(u)}-\frac{5}{A(u)^2}+6\right\}
\label{Def::I404}
\,,
 \end{align}
 \end{subequations}
where $\mu$ is the reduced mass and $A(u) := 1-e_t \cos u$.
The above multipole moments contain terms with
certain equal
structure. Evaluating the exponential functions
$e^{-i\magnetic\phi}$ with $\magnetic=1,2,3,4$ to $1$pN accuracy, one
obtains factors of the form $A(u)^{-n}$ and 
$\sin(u)/A(u)^n$.
\footnote{
$\sin u$ and $\cos u$ of higher power than one in the denominator
can always be expressed again in $A(u)$ and $\sin u$.
}
Those terms can be expanded first
in $\sin$ and $\cos$ series of $u$ and afterwards in
terms of $\Mean$ with the help
of Eqs. (\ref{Eq::sinmE}) and (\ref{Eq::cosmE}).
We will calculate the expansion coefficients in the next section.
Some calculation steps will be required repeatedly and we will
collect them as ingredients for a frequency-domain GW recipe.
\section{Ingredients for Fourier decomposition}
\label{Sec::Ingredients}
The Fourier-domain GW form requires computation of relatively simple structures
which will combine in the tensor-spherical harmonics. Let us start with the
most fundamental quantities which we compute from the scratch on the
one hand and some which we collect from the literature on the other.

\begin{enumerate}
 \item The inverse scaled radial separation with an arbitrary integer exponent $n>0$:
\begin{eqnarray}
\label{Eq::Aexpn}
 \frac{1}{(1-e \cos u)^n}
					&=&  1 + b^{(n)}_0 + \sum_{j=1}^{\infty} b^{(n)}_j \cos ju \,, \\
	b^{(n)}_{0}		&=&  \sum_{i=1}^{\infty} \beta^{(n)}_{2i,i} \,,							\label{Eq::Def_b_n_0}	\\
	b^{(n)}_{j>0} 	&=&  \sum_{i=0}^{\infty} \beta^{(n)}_{j+2i,i} + \beta^{(n)}_{j+2i, j+i} \,, 	\label{Eq::Def_b_n_j>0}	\\
	\beta^{(n)}_{m,k} &:=& \frac{(n+m-1)!}{{(n-1)!}} \,\frac{1}{m!} \, \frac{e^m}{2^m} \binom{m}{k} \nonumber \\
					& =& e^m \, \frac{1}{2^m} \, \binom{m}{k}\,\frac{1}{m!} \, \prod_{i=0}^{m-1} (n + i) 
\,.
\end{eqnarray}
Equation (\ref{Eq::Aexpn}) is proven in Appendix \ref{App:Proof}. Note that from here onwards,
$e$ can be set $e_{\phi}$ or $e_t$, depending on the context.

 \item	Trigonometric functions of multiples of the eccentric anomaly $u$
	in terms of Bessel functions (see \cite{Watson:1980}, p. 555),
\begin{eqnarray}
 \sin {m u}	&=& m \sum_{n=1}^{\infty} \frac{1}{n} \left( J_{n-m}(n e_t) + J_{n+m}(n e_t) \right)	\sin (n \Mean)		
= \sum_{n=1}^{\infty} \SKE{m}{n} \sin {n\Mean} \,,\label{Eq::sinmE}
\\
\SKE{m}{n} &:=& \frac{m}{n} \left( J_{n-m}(n e_t) + J_{n+m}(n e_t) \right)\,,
\\
 \cos {m u}	&=& m \sum_{n=1}^{\infty} \frac{1}{n} \left( J_{n-m}(n e_t) -  J_{n+m}(n e_t) \right)	\cos (n \Mean)
= \sum_{n=1}^{\infty} \CKE{m}{n} \cos {n \Mean}\,,  \hspace{1cm} {\rm (for~~} m \ge 1)
\label{Eq::cosmE}
\\
\cos u &=& -\frac{1}{2} e_t + 2 \sum_{n=1}^{\infty} J'_{n} (n e_t) \cos (n\Mean) \,,
\\
\CKE{m}{n}	&:=& \frac{m}{n} \left( J_{n-m}(n e_t) -  J_{n+m}(n e_t) \right) -\frac{e_t}{2} \delta_{m1} \delta_{n0}\,. \label{Eq::CoeffCos_mE}
\\
J_{n}(y)		&=& \frac{1}{\pi} \, \int_{0}^{\pi} \cos \left( n x -  y \sin x \right) {\rm d}x
\,.
\end{eqnarray}
Note that the above Eqs. (\ref{Eq::sinmE}) - (\ref{Eq::CoeffCos_mE}) are only valid taking $e_t$ because of the structure of the KE.

 \item The decomposition of $v$ in $\Mean$  (see \cite{Colwell:1993}, p. 33),
\begin{equation}
\label{Eq::vSumInM}
 v_{(e=e_t)}=\Mean + \sum_{m=1}^{\infty} G_m(e_t) \, \sin m \Mean \,,
\end{equation}
with $G_m$ defined as
\begin{equation}
\label{Def::G_in_J}
 G_m(e) = \frac{2}{m} J_m (m e) + \sum_{s=1}^{\infty} \alpha^s \left[ J_{m-s}(me) - J_{m+s}(me)\right] \,,
\end{equation}
and $\alpha$ extractable from
\begin{equation}
\label{Def::alpha}
 e = \frac{2\alpha}{1+\alpha^2}\,.
\end{equation}

 \item The decomposition of the 1pN accurate Kepler equation itself in terms of the mean anomaly (see \cite{Watson:1980}, p. 553),
\begin{equation}
u = \Mean + \sum_{n=1}^{\infty} \frac{2}{n} J_n (n e_t) \sin (n \Mean)	\,,	
\end{equation}
\end{enumerate}
which converges for all $e_t<1$.
These inputs are sufficient to obtain a 1pN-accurate F-domain decomposition
of the 1pN far-zone GW field including 1pN-accurate orbital dynamics. More
complicated expressions will be based on those above and can be computed
more or less laboriously. A detailed appendix of this article will give some
proves for resummation rules for these terms.
The next section will deal with certain exponential functions including
the 1pN-accurate phase. They will be combined with the fast oscillatory
terms
in $f_{l \magnetic}$ and $g_{l \magnetic}$
and have individual Fourier decompositions
as shown below.
\section{Decomposition of the elements appearing in the multipoles}
\label{Sec::decomp_elements}

This section is subject to the Fourier decomposition of
single terms in the multipole field expressions. We will
apply the aforementioned steps to obtain the individual
Fourier decompositions.
The ``magnetic'' number (the factor of $-i\, \phi$ in the phase exponentials of Eqs. (\ref{Def::I222}) - (\ref{Def::I404}) ) which we shall call $\magnetic$
in the multipoles dictates terms of the form
$e^{-i \magnetic \phi}$, which have to be expanded in
$\epsilon$.
\begin{eqnarray}
\label{Eq:ExpInphi}
v_{e_t} &:=& 2\, \arctan \left [ {\sqrt{\frac{1+e_{t}}{1-e_{t}}} \, \tan \frac{u}{2}} \right ] \label{Def::vet} \,,	\\
v_{e_\phi} &:=& 2\, \arctan \left [ {\sqrt{\frac{1+e_{\phi}}{1-e_{\phi}}} \, \tan \frac{u}{2}} \right ] \label{Def::vephi} \,,	\\
 e^{-i \, \magnetic \, \phi} &=&
					e^{-i\magnetic \phi_0} \, e^{-i\magnetic(1+\peri) v_{e_\phi}}
				=	e^{-i\magnetic \phi_0} \, e^{-i\magnetic v_{e_\phi}}\,e^{-i\magnetic \peri v_{e_t}}
\nonumber \,, \\
				&=& 
					 e^{-i\magnetic \phi_0} \, 
					\left( \frac{\cos u - e_\phi}{1-e_\phi \cos u}
					-i \, \sqrt{1-e_\phi^2}\, \frac{\sin u}{1-e_\phi cos u}\right)^\magnetic \, e^{-i\magnetic  \peri v_{e_t}}	\nonumber \\
				&=&
					e^{-i\magnetic \phi_0}
					 \left( \frac{\cos u - e_\phi}{1-e_\phi \cos u}
					-i \, \sqrt{1-e_\phi^2}\, \frac{\sin u}{1-e_\phi cos u}\right)^\magnetic
					{\rm exp} \left[-i \magnetic \peri \left( \Mean + \sum_{j=1}^{\infty} G_j(e_t) \sin j \Mean \right) \right] \nonumber \\
				&=&
					e^{-i\magnetic \phi_0}
					 \left( \frac{\cos u - e_\phi}{1-e_\phi \cos u}
					-i \, \sqrt{1-e_\phi^2}\, \frac{\sin u}{1-e_\phi \cos u}\right)^\magnetic
					\times
\nonumber \\ && 
					 {\rm exp}\left[-i \magnetic \peri  \Mean \right] \, \left(1 -i \magnetic \peri \, \sum_{j=1}^{\infty} G_j(e_t) \sin j\Mean \right)
\,.
\end{eqnarray}
For the first line we took the definition of $v$ as a function
of $u$ and $e_t$, Eq. (\ref{eq::vparm}), and Eq. (\ref{eq::angleparm}).
In the last line above we truncated the exponential of the sin series after the linear order in $\peri$, because $\peri$ is already of 1pN order
and used the Fourier-domain expansion Eq. (\ref{Eq::vSumInM}) for $v$.
We are allowed to do this because there is no secular contribution coming up in the exponent due to its periodicity of $2\pi$ in $\Mean$
and its average to zero in the interval $[0, 2\pi]$.

In the multipole moments the numbers $\magnetic=(0,1,2,3,4)$ appear,
so we preserve all contributions with these $\magnetic$ up to the order of $\epsilon$
required.
%
%
\begin{eqnarray}
e^{-i v_{e_\phi}}
&=&
\frac{1}{e_t}
\Biggl\{
\frac{i \OTS e_t \sin u-e_t^2+1}{A(u)}-1
 \Biggr\}
+{\cal O} (\epsilon^2)
\,,
\\
e^{-i 2 v_{e_\phi}}
&=&
\frac{1}{e_t^2}
\Biggl\{
2-e_t^2
+  \frac{4 \left(e_t^2-1\right)
+2 i e_t \OTS \sin u
}{A(u)}+\frac{
2  \left(e_t^2-1\right){}^2
-{2 i e_t \OTS^3
 \sin u }}{A(u)^2}
\nonumber \\ && 
+\epsilon^2
{\En} \Biggl[
8 (\eta -4)
+\frac{8 (\eta -4) \left(e_t^2-3\right)-\frac{4 i
   (\eta -4) e_t \left(e_t^2-2\right) \sin
   (u)}{{\OTS}}}{A(u)}+\frac{-24 (\eta -4)
   \left(e_t^2-1\right)+\frac{16 i (\eta -4) e_t \left(e_t^2-1\right)
   \sin (u)}{{\OTS}}}{A(u)^2}
\nonumber \\ &&
+\frac{-8 (\eta -4)
   \left(e_t^2-1\right){}^2-8 i {\OTS} (\eta -4) e_t
   \left(e_t^2-1\right) \sin (u)}{A(u)^3}\Biggr]
+{\cal O}(\epsilon^4)
\,,
\\
e^{-i 3 v_{e_\phi}}
&=&
\frac{1}{e_t^3}
\Biggl\{
   3 e_t^2-4
+ \frac{3 \left(e_t^4-5 e_t^2+4\right)-\frac{i e_t \left(e_t^4-5 e_t^2+4\right) \sin u}{\OTS}}{A(u)}
+ \frac{-12 \left(e_t^2-1\right){}^2
+{8 i e_t \OTS^3
 \sin u}}{A(u)^2}
\nonumber \\ && 
+ \frac{
-4 \left(e_t^2-1\right){}^3
-{4 i e_t \OTS^5   \sin u}}{A(u)^3}
\Biggr\}
+{\cal O} (\epsilon^2)
\,,
\\
e^{-i 4 v_{e_\phi}}
&=&
\frac{1}{e_t^4}
\Biggl\{
  e_t^4-8 e_t^2+8
+\frac{-16 \left(e_t^4-3e_t^2+2\right)-4 i e_t \sqrt{1-e_t^2} \left(e_t^2-2\right) \sin u}{A(u)}
\nonumber \\ && 
+\frac{-8 \left(e_t^2-6\right) \left(e_t^2-1\right){}^2 - 4 i e_t \sqrt{1-e_t^2} \left(e_t^4-7 e_t^2+6\right) \sin u}{A(u)^2}
\nonumber \\ && 
+\frac{32 \left(e_t^2-1\right){}^3+24 i e_t \left(1-e_t^2\right){}^{5/2} \sin u}{A(u)^3}
\nonumber \\ && 
+\frac{8 \left(e_t^2-1\right){}^4 - 8 i e_t \left(1-e_t^2\right){}^{7/2} \sin  (u)}{A(u)^4}
\Biggr\}
+{\cal O} (\epsilon^2)
\,.
\end{eqnarray}
Those terms will mix with $ f_{l\magnetic}$ and  $g_{l\magnetic}$.
Let us get more precise now and start decomposing
the $A(u)^{-n}$ term. With the help of Eq. (\ref{Eq::Aexpn})
we get
\begin{eqnarray}
\label{Eq::Decomp_Au_Cos_ju}
 \frac{1}{(1-e_t \, \cos u)^n}
	&=& 1 + \sum_{m=1}^{\infty} \sum_{k=0}^{m} \beta^{(n)}_{m, k} \cos [(m-2k) u] \nonumber  \\
	&=& 1 + b^{(n)}_0 + \sum_{k=1}^{\infty} b^{(n)}_{k} \, \cos k u \,,
\end{eqnarray}
where we have collected those terms with the same {\em positive}
frequency. Appendix \ref{App:Proof} will give deeper explanation how to select
the frequencies with the same absolute value.
The same contribution -- being multiplied with $\sin u$ --
will be decomposed analogously, but having $\sin m u$ this time
as an odd function of $u$.
\begin{eqnarray}
\label{Eq::Decomp_SinuAu_Sin_ju}
  \frac{\sin u}{(1-e_t \, \cos u)^n}
	&=& \left( 1 + b^{(n)}_0 +  \sum_{j=1}^{\infty} {b}^{(n)}_j \, \cos ju
		\right) \, \sin u \nonumber \\
	&=& \left( 1 + \sum_{k=1}^{\infty} \beta^{(n)}_{2k,k} \right) \sin u
		+ \frac{1}{2} \left( \sum_{m=2}^{\infty} 
						\left[ b^{(n)}_{m-1} - b^{(n)}_{m+1} \right] \sin mu
	\right)
 \nonumber \\
	&=& \sum_{j=1}^{\infty} \SA{n}{j} \, \sin {j u} \,, 
\\
\SA{n}{1}
	&:=& 1 + \sum_{k=1}^{\infty} \beta^{(n)}_{2k,k} - \frac{1}{2} b^{(n)}_2 \,, 		\label{Def::SA_1} \\
\SA{n}{j>1}
	&:=& \frac{1}{2} \left( b^{(n)}_{j-1}  - b^{(n)}_{j+1} \right)						\label{Def::SAj>1}
\,.
\end{eqnarray}

Remembering Eqs. (\ref{Eq::sinmE}) and (\ref{Eq::cosmE}), we get

\begin{eqnarray}
\frac{1}{(1-e_t \cos u)^n}	&=&
1+ b^{(n)}_0 + \sum_{k=1}^{\infty} b^{(n)}_k \cos k u
=	1+ b^{(n)}_0 + \sum_{k=1}^{\infty} b^{(n)}_{k} \left( \sum_{j=1}^{\infty} \CKE{k}{j} \cos j\Mean \right)
\nonumber \\
&=& 1+ b^{(n)}_0 + \sum_{j=1}^{\infty} \left( \sum_{k=1}^{\infty} \CKE{k}{j} b^{(n)}_{k} \right) \cos j\Mean
\nonumber \\
&=& 1+ b^{(n)}_0 +  \sum_{n=1}^{\infty} \AKE{n}{j} \cos j\Mean
\label{Eq::Au_KE}
\,,
\\
\frac{\sin u}{(1-e_t \cos u)^n}	&=&
 \sum_{k=1}^{\infty} \SA{n}{k} \sin k u = \sum_{k=1}^{\infty} \SA{n}{k} \left( \sum_{j=1}^{\infty} \SKE{k}{j} \sin j\Mean \right)
= \sum_{j=1}^{\infty} \left( \sum_{k=1}^{\infty} \SKE{k}{j} \SA{n}{k} \right) \cos j\Mean
\nonumber \\
&=& \sum_{j=1}^{\infty} \SAKE{n}{j} \sin j\Mean
 \label{Eq::Sinu_Au_KE}
\,,
\\
\AKE{n}{0}	&:=&	1 + b^{(n)}_{0} \,, \\
\AKE{n}{j>0}	&:=&	\left( \sum_{m=1}^{\infty} \CKE{m}{j} b^{(n)}_{m} \right)
\,,
\\
\SAKE{n}{j>0}	&:=& \left( \sum_{m=1}^{\infty} \SKE{m}{j} \SA{n}{m} \right)
\,.
\end{eqnarray}
%
The abbreviations {$\AKE{n}{j}$} and $\SAKE{n}{j}$
stand for the j$^{\rm th}$ contribution of $A(u)^{-n}$
and $ A(u)^{-n} \, \sin u$. 
Of course, both expressions will not take their place alone,
but will be multiplied with the expansion of the $e^{in\peri v_{e_t}}$ term
due to Eq. (\ref{Eq:ExpInphi}).
We Fourier decompose the following products of series as they appear in the field multipole moments.
The first quantity consists of two pure sin series.
Factor 1 is the expansion of $\sin(u)\,A(u)^{-n}$ in $\Mean$ and factor 2 is the non-secular part of the exponential of the $v(l)$-term.
\begin{eqnarray}
 \left(
\sum_{k=1}^{\infty}
	\SAKE{n}{k} \sin k \Mean
\right)
\left(
\sum_{m=1}^{\infty}
	\vKE{m} \sin m \Mean
\right)
&=&\sum_{k=1}^{\infty} \sum_{m=1}^{\infty}
\frac{1}{2} 
\SAKE{n}{k} \vKE{m}
\left(
\cos[k-m] \Mean - \cos[k+m] \Mean
\right)
\nonumber\\
&=& \sum_{j=0}^{\infty} \ProdSS{j} \cos j \Mean
\,.
\end{eqnarray}
The coefficients  $\ProdSS{j}$ (``SS'' stands for the product of two sin series) are defined as follows.
\begin{eqnarray}
\ProdSS{0} &:=& \frac{1}{2}
\sum_{k=1}^{\infty} \SAKE{n}{k} \vKE{k} \,, \\
\ProdSS{j} &:=& \frac{1}{2} \left\{
\sum_{k=1}^{\infty}
\left(
 \SAKE{n}{k} \vKE{k+j} +  \underbrace{\SAKE{n}{k} \vKE{k-j}}_{{\rm for ~} k>j}
\right)
-
\underbrace{
\sum_{k=1}^{j-1} \SAKE{n}{k} \vKE{j-k}
}_{{\rm for ~} j>1}
\right\}
\,.
\label{Eq::DoubleSum_Sin_Sin}
\end{eqnarray}

The second quantity consists of a $\sin$ and a $\cos$ series.
Factor 1 is the decomposition of $A(u)^{-n}$ in $\Mean$ and
factor 2 is -- again -- the non-secular part of the exponential
of the $v(l)$-term.
\begin{eqnarray}
 \left(
\sum_{k=1}^{\infty} \AKE{n}{k} \cos {k \Mean}
\right)
 \left(
\sum_{m=1}^{\infty} \vKE{m} \sin{m \Mean}
\right)
&=&
\frac{1}{2}
\sum_{k=1}^{\infty}
\sum_{m=1}^{\infty}
 \AKE{n}{k} \,  \vKE{m}
\left(
\sin [k+m] \Mean
-\sin [k-m] \Mean
\right)
\nonumber \\
= \sum_{j=1}^{\infty} \ProdCS{j} \sin {j \Mean}
\,.
\label{Eq::DoubleSum_Sin_Cos}
\end{eqnarray}
As well as for $\ProdSS{j}$, we provide the definition of the $\ProdCS{j}$ as combinations
of the elements of the two series.
\begin{eqnarray}
 \ProdCS{j} :=
\frac{1}{2×}
\left\{
\underbrace{
  \sum_{k=1}^{j-1}	\left( \AKE{n}{k} \vKE{j-k} \right)
}_{{\rm for~} j>1}
- \sum_{k=1}^{\infty}	\left( \underbrace{\AKE{n}{k} \vKE{k-j}}_{{\rm for~}k>j}-\AKE{n}{k} \vKE{k+j} \right)
\right\}
\,.
\end{eqnarray}
The two above summation formulas will also be proven in
Appendix \ref{App::ProdSinCos} and \ref{App::ProdSinSin}.
Please keep in mind that the $(n)$ is always present in those terms and reminds
the user of keeping the right exponential of $A(u)^{-n}$ and that those functions
always depend on the value of $e$.
Additionally note that the statements ``${\rm for~} j>1$'' and ``${\rm for~} k>j$''
ensure that, for example,
the first term in (4.19) is not able to contribute to the first harmonic in $l$ and that
the indices of the decomposition coefficients do not become zero, or else, these elements
are set to zero automatically.

The next step is to sum up and summarise the terms appearing in the multipole moments.
Using Eq. (\ref{Eq::vSumInM}), they also contain absolute static elements and will be listed below.

\begin{eqnarray}
\frac{1}{(1-e \cos u)^n}\, e^{ (-iq\peri) v_{e_t}}
&=&	\Biggl\{ (1 + b^{(n)}_0) + \sum_{j=1}^{\infty} \AKE{n}{j} 
														 \cos j \Mean \Biggr\}
		\Biggl\{ 1 +  (-i q \peri) \sum_{m=1}^{\infty} \vKE{m} \sin m\Mean \Biggr\}
		\Biggl\{ e^{(-i q \peri) \Mean} \Biggr\}
		\nonumber \\
&=&\Biggl\{
		\Biggl[ 1 +  (-i q \peri) \sum_{j=1}^{\infty} \vKE{j} \sin j\Mean \Biggr]\left( 1 + b^{(n)}_0 \right)
	+	\Biggl[ \sum_{j=1}^{\infty} \AKE{n}{j} 
									 \cos j\Mean \Biggr] \nonumber \\
&&	+	(-iq\peri) \sum_{j=1}^{\infty} \ProdCS{j} \sin j\Mean
	\Biggr\}
	\Biggl\{ e^{-(iq\peri)\Mean} \Biggr\} \nonumber \\
&=&
\Biggl\{	   \left( 1+b^{(n)}_{0} \right)
		+  (-iq\peri) \sum_{j=1}^{\infty}\left[ (1+b^{(n)}_{0}) \vKE{j}
		+ \ProdCS{j} \right] \sin{j\Mean}
 \nonumber \\ &&
		+  \sum_{j=1}^{\infty} \AKE{n}{j} 
									\cos j\Mean
\Biggr\} \Biggl\{ e^{-(iq\peri)\Mean} \Biggr\}
\,.
\end{eqnarray}
We will perform the equivalent steps for
\begin{eqnarray}
 \frac{\sin u}{(1-e \cos u)^n}\, e^{ (-iq\peri) v_{e_t}}
&=& \Biggl\{ \sum_{j=1}^{\infty} \SAKE{n}{j} \sin j \Mean \Biggr\}
		\Biggl\{ 1 +  (-i q \peri) \sum_{m=1}^{\infty} \vKE{m} \sin m\Mean \Biggr\}
		\Biggl\{ e^{(-i q \peri) \Mean} \Biggr\}
		\nonumber \\
&=& \Biggl\{
	\sum_{j=1}^{\infty} \SAKE{n}{j} \sin j \Mean
+	(-iq\peri) \sum_{j=0}^{\infty} \ProdSS{j} \cos j\Mean
\Biggr\}
\Biggl\{ e^{(-i q \peri) \Mean} \Biggr\}
\,.
\end{eqnarray}
This will be shortened by writing
\begin{eqnarray}
 \frac{1}{(1-e \cos u)^n}\, e^{ (-iq\peri) v_{e_t}}
&=&
\left\{
\funcAKEc{n,q}_0
+ \sum_{j=1}^{\infty} 	\funcAKEc{n,q}_ j  \cos j \Mean
+ \sum_{j=1}^{\infty} 	\funcAKEs{n,q}_ j  \sin j \Mean
\right\}
e^{-iq \peri \Mean}
\,,	
\label{Eq::Separate_Ainvnq_M}
\\
 \frac{\sin u}{(1-e \cos u)^n}\, e^{ (-iq\peri) v_{e_t}}
&=&
\left\{
\funcSinAKEc{n,q}_0
+ \sum_{j=1}^{\infty} 	\funcSinAKEc{n,q}_ j  \cos j \Mean
+ \sum_{j=1}^{\infty} 	\funcSinAKEs{n,q}_ j  \sin j \Mean
\right\}
e^{-iq \peri \Mean}
 \,,	
\label{Eq::Separate_SinAinvnq_M}
\\
\funcAKEs{n,q}_{j>0}			&:=&	(-i q \peri) \left[ (1+b^{(n)}_{0}) \vKE{j} +  \ProdCS{j} \right]	\,,	\\
\funcAKEc{n,q}_{0}			&:=&	 \left[(1+b^{(n)}_{0})		 					  \right]	\,,	\\
\funcAKEc{n,q}_{j>0}			&:=&	\AKE{n}{j}\,, \\
\funcSinAKEc{n,q}_{j>0}		&:=&	(-i q \peri) \ProdSS{j}	\,,	\\
\funcSinAKEs{n,q}_{j>0}		&:=&	\SAKE{n}{j}	\,.
\end{eqnarray}

We are now in the lucky position to decompose all point particle contributions
to the radiation field amplitude in terms of the above defined
$\funcAKEs{n,q}_j$,
$\funcAKEc{n,q}_j$,
$\funcSinAKEs{n,q}_j$, and
$\funcSinAKEc{n,q}_j$.
For simplicity, we introduce
\begin{eqnarray}
 \frac{1}{(1-e \cos u)^n} e^{-iq\peri v_{e_{t}}} &=:&\funcAnq{n}{q}	\, e^{-iq\peri\Mean} \,,
\label{Def::FuncNQ}	\\
 \frac{\sin u}{(1-e \cos u)^n} e^{-iq\peri v_{e_{t}}}&=:&\funcSinAnq{n}{q}		\, e^{-iq\peri\Mean}	\,,
\label{Def::FuncSinNQ}
\end{eqnarray}
and use them from now on in the multipoles, where $n$ and $q$ will also be allowed to be 0.
We finally rewrite our wave forms in the following manner,
\begin{eqnarray}
 e^{-i \magnetic \phi} f_{\magnetic l}
&=& e^{-i\magnetic \phi_0} e^{-i \magnetic v_{e_{\phi}}} e^{-i\magnetic \peri v_{e_t}}f_{\magnetic l}  \nonumber \\
&=& e^{-i\magnetic \phi_0} \left[e^{-i \magnetic v_{e_{t}}} \, f_{\magnetic l}  + {\cal O}( \epsilon^2) \right] \, e^{-i\magnetic \peri v_{e_t}} \nonumber \\
&=& e^{-i\magnetic \phi_0} \left[e^{-i \magnetic v_{e_{t}}} \, f_{\magnetic l}  + {\cal O}( \epsilon^2) \right] \, e^{-i\magnetic \peri \Mean}
		\left( 1+(-i\magnetic\peri) \sum_{j=1}^{\infty}G_j \sin j\Mean)\right) \nonumber \\
&=& e^{-i\magnetic \phi_0}
\sum_{j}^{}
 \left[
 \kappa_{[j\magnetic]} \, \funcAnq{j}{\magnetic} + \tilde \kappa_{[j\magnetic]} \, \funcSinAnq{j}{\magnetic}
	 \right] \, e^{-i\magnetic \peri \Mean}
\,,
\end{eqnarray}
with some $\kappa_{[j\magnetic]}$ and $\tilde \kappa_{[j\magnetic]}$ to be determined.

\section{Decomposition of the multipole moments}
\label{Sec::Fou_Decomp_Multipoles}
We will decompose the multipoles in terms of the
functions $\funcAnq{n}{q}$ introduced above next.
These functions will have prefactors
$\alpha^{(\magnetic)}_{n}$, $\beta^{(\magnetic)}_{n}$, $\dots$
and
$\salpha^{(\magnetic)}_{n}$, $\sbeta^{(\magnetic)}_{n}$, $\dots$
for the different multipole types, where
$\magnetic$ denotes again the magnetic
number and $n$ is the same exponent label as in Defs.
(\ref{Def::FuncNQ}). and (\ref{Def::FuncSinNQ}).
\begin{eqnarray}
\stackrel{(2)}{I}_{22}
&=& 
{
8 \sqrt{\frac{2 \pi }{5}} \,  E \, \mu  \, e_t^{-2} \,
 e^{-2i\peri \Mean}
 e^{-i2\phi_0}
\left\{
\sum_{k=0}^{5}  \alpha_{[k2]} \funcAnq{k}{2}
+\sum_{k=1}^{5}  \salpha_{[k2]} \funcSinAnq{k}{2}
\right\}
}\,,
\\
	\alpha_{[0 2]}	&~:=~&	\left[ 2-e_t^2 \right]
	 + \epsilon^2 \,  \frac{1}{14} {\En} \left[(9 \eta -3) e_t^2+94 \eta -442\right]  \,, 
\\
	\alpha_{[1 2]}	&~:=~&	\left[ e_t^2-2 \right] 
	+ \epsilon^2 
	\frac{1}{14} {\En} \left[(285 \eta -781) e_t^2-346 \eta +1450\right] \,, 
\\
	\alpha_{[2 2]}	&~:=~&	
		2 \left(e_t^2-1\right)
		+\epsilon^2\,{\En}	\frac{1}{21} \left[84 (3 \eta -8) e_t^4+(3713-1458 \eta ) e_t^2+981 \eta -2861\right] \,, \\ 
	\alpha_{[3 2]}	&~:=~&
		2 \left[e_t^2-1\right]^2
+\epsilon^2 {\En}
		\biggl[
		-\frac{1}{21} \left(e_t^2-1\right) \left((663 \eta -1537) e_t^2-1005 \eta +1861\right)
		\biggr] 
\,, \\
	\alpha_{[4 2]}	&~:=~&
			\epsilon^2 {\En} \left[ \frac{2}{21} (219 \eta -262) \left(e_t^2-1\right)^2) \right] 
\,, \\
	\alpha_{[5 2]}	&~:=~&	
		\epsilon^2 \En \left[	\frac{4}{7} (3 \eta -1) \left(e_t^2-1\right)^3
					\right] 
\,, \\
	\salpha_{[1 2]}	&~:=~&
					\left[ 2 i e_t \sqrt{1-e_t^2}  \right] + \epsilon^2 \left[	\frac{ i {\En} e_t \left((109-19 \eta ) e_t^2+47 \eta -221\right)}{7 \sqrt{1-e_t^2}} \right]
\,, \\
	\salpha_{[2 2]}	&~:=~& 
					\epsilon^2 \left[ - i \frac{2 {\En} e_t \left(2 (3 \eta -8) e_t^4+(46-15 \eta )  e_t^2+9 (\eta -4)\right)}{\sqrt{1-e_t^2}} \right]
\,, \\
	\salpha_{[3 2]}	&~:=~&
					\left[-2 i e_t \left(1-e_t^2\right)^{3/2} \right]
					+ i \epsilon^2 \biggl[ \frac{1}{21} {\En} e_t \sqrt{1-e_t^2} \left((1531-645 \eta )  e_t^2+603 \eta -1349\right) \biggr]
\,, \\
	\salpha_{[4 2]}	&~:=~&	i \epsilon^2 \biggl[ -\frac{2}{21} {\En} (201 \eta -256) e_t   \left(1-e_t^2\right)^{3/2} \biggr]
\,, \\
	\salpha_{[5 2]}	&~:=~& i \epsilon^2  \left[ \frac{4}{7} {\En} (3 \eta -1) e_t \left(1-e_t^2\right){}^{5/2} \right]
\,, \\
\hline
\stackrel{(2)}{I}_{21} &~=~& 0 \,, \\
\hline
\stackrel{(2)}{I}_{20} &~=~& 
-16 \sqrt{\frac{\pi }{15}} E \mu
\Bigl\{
	\alpha_{[0 0]} \, \funcAnq{0}{0}
+	\alpha_{[1 0]} \, \funcAnq{1}{0}
+	\alpha_{[2 0]} \, \funcAnq{2}{0}
+	\alpha_{[3 0]} \, \funcAnq{3}{0}
\Bigr\} \,, \\
	\alpha_{[0 0]} &~:=~&	1-\frac{3}{14}	{\En} \epsilon ^2 (3 \eta -1) \,, \\		
	\alpha_{[1 0]} &~:=~&	\frac{1}{14}		{\En} \epsilon ^2 (51 \eta -115)-1	\,,	\\
	\alpha_{[2 0]} &~:=~&	\frac{1}{7}		{\En} \epsilon ^2 (4-19 \eta )	\,,	\\	
	\alpha_{[3 0]} &~:=~&	\frac{2}{7}		{\En} \epsilon ^2 (\eta -26) \left(e_t^2-1\right)	\,,	\\ 
\hline
\stackrel{(2)}{S}_{22} &~=~& 0 \,,	\\
\hline
\stackrel{(2)}{S}_{21} &~=~&
\frac{32}{3} \sqrt{\frac{\pi }{5}} \sqrt{1-{e_t}^2} \eta  { \mEn }^{3/2} ({m_1}-{m_2}) \,
e^{-i\peri \Mean}
e^{-i\phi_0} \times
\nonumber \\ && 
\Bigl\{
	\beta_{[2 1]} \funcAnq{2}{1}
+	\beta_{[3 1]} \funcAnq{3}{1}
+	\sbeta_{[3 1]} \funcSinAnq{3}{1}
\Bigr\}
 \,,	\\
	\beta_{[2 1]}	&~:=~& -\frac{1}{e_t}	\,,	\\
	\beta_{[3 1]}	&~:=~& \frac{1}{e_t}-e_t	\,,	\\
	\sbeta_{[3 1]}	&~:=~&	- i {\OTS}		\,,	\\
\hline
\stackrel{(2)}{S}_{20}	&~=~&	0	\,,	\\
\hline
 \stackrel{(3)}{I}_{33} 
&~=~& 
{
{8\left(\frac{2\pi}{21} \right)^{\frac{1}{2}}	\eta	(m_1-m_2) (-E)^{3/2}
e^{-i3\peri \Mean}
}
e^{-i3\phi_0} \,
\left\{
\sum_{k=0}^{5}  \gamma_{[k3]} \funcAnq{k}{3}
+\sum_{k=1}^{5}  \sgamma_{[k3]} \funcSinAnq{k}{3}
\right\}
}
\,,
\\
	\gamma_{[0 3]}  &~:=~&	\frac{i \sqrt{1-e_t^2} \left(e_t^2-4\right)}{e_t^3} \,, \\ 
	\gamma_{[1 3]}  &~:=~&	-\frac{i \sqrt{1-e_t^2} \left(e_t^2-4\right)}{e_t^3}\,, \\
	\gamma_{[2 3]}  &~:=~&	\frac{i \left(12-7 e_t^2\right) \sqrt{1-e_t^2}}{2 e_t^3} \,, \\
	\gamma_{[3 3]}  &~:=~&	\frac{i \sqrt{1-e_t^2} \left(e_t^2-1\right) \left(e_t^2+4\right)}{2 e_t^3}\,, \\
	\gamma_{[4 3]}  &~:=~&	-\frac{10 i \sqrt{1-e_t^2} \left(e_t^2-1\right){}^2}{e_t^3} \,, \\
	\gamma_{[5 3]}  &~:=~&	-\frac{6 i \sqrt{1-e_t^2} \left(e_t^2-1\right){}^3}{e_t^3} \,, \\
	\sgamma_{[1 3]}  &~:=~&	\frac{4}{e_t^2}-3 \,, \\
	\sgamma_{[2 3]}  &~:=~&	0 \,, \\
	\sgamma_{[3 3]}  &~:=~&	-\frac{5 e_t^2}{2}-\frac{6}{e_t^2}+\frac{17}{2} \,, \\
	\sgamma_{[4 3]}  &~:=~&	-\frac{4 \left(e_t^2-1\right){}^2}{e_t^2}\,, \\
	\sgamma_{[5 3]}  &~:=~&	-\frac{6 \left(e_t^2-1\right){}^3}{e_t^2} \,, \\
\hline
\stackrel{(3)}{I}_{32} &~=~& 0 \,, \\
\hline
\stackrel{(3)}{I}_{31} &~=~&
{
e^{-i\,\peri \Mean} \,
e^{-i\,\phi_0}
8 \sqrt{\frac{2 \pi }{35}} (-E)^{3/2} \left(m_1-m_2\right) \eta \times
\left\{
\sum_{k=0}^{3}		\gamma_{[k1]}	\funcAnq{k}{1}
+\sum_{k=1}^{3}		\sgamma_{[k1]}	\funcSinAnq{k}{1}
\right\}
}\,,
\\
	\gamma_{[0 1]} &~:=~&	\frac{i \sqrt{1-e_t^2}}{e_t} \,, \\ 
	\gamma_{[1 1]} &~:=~&	-\frac{i \sqrt{1-e_t^2}}{e_t} \,, \\ 
	\gamma_{[2 1]} &~:=~&	-\frac{5 i \sqrt{1-e_t^2}}{6 e_t} \,, \\ 
	\gamma_{[3 1]} &~:=~&	-\frac{5 i \sqrt{1-e_t^2} \left(e_t^2-1\right)}{6 e_t} \,, \\ 
	\sgamma_{[1 1]} &~:=~&	-1 \,, \\ 
	\sgamma_{[2 1]} &~:=~&	0 \,, \\ 
	\sgamma_{[3 1]} &~:=~&	-\frac{5}{6} \left(e_t^2-1\right) \,, \\ 
\hline
\stackrel{(3)}{I}_{30}	&~=~& 0\,, \\
\hline
\stackrel{(3)}{S}_{33}	&~=~& 0\,, \\
\hline
\stackrel{(3)}{S}_{32}	&~=~&
{
e^{-i \,2\,\peri \Mean}\,
e^{-i \,2\phi_0}
\frac{8}{3} \sqrt{\frac{2 \pi }{7}} E^2 \sqrt{1-e_t^2} \mu  (1-3 \eta ) \times
\left\{
	\sum_{k=2}^{5}		\delta_{[k2]}	\funcAnq{k}{2}
+	\sum_{k=3}^{5}		\sdelta_{[k2]}	\funcSinAnq{k}{2}
\right\}
} \,,
\\
	 \delta_{[2 2]}  &~:=~&	-\frac{2 i \sqrt{1-e_t^2}}{e_t^2} \,, \\ 
	 \delta_{[3 2]}  &~:=~&	\left(2 i-\frac{2 i}{e_t^2}\right) \sqrt{1-e_t^2} \,, \\ 
	 \delta_{[4 2]}  &~:=~&	\left(\frac{10 i}{e_t^2}-10 i\right) \sqrt{1-e_t^2} \,, \\ 
	 \delta_{[5 2]}  &~:=~&	-\frac{6 i \sqrt{1-e_t^2} \left(e_t^2-1\right){}^2}{e_t^2} \,, \\ 
	\sdelta_{[3 2]}  &~:=~&	\frac{2}{e_t}-e_t \,, \\ 
	\sdelta_{[4 2]}  &~:=~&	\frac{4}{e_t}-4 e_t \,, \\ 
	\sdelta_{[5 2]}  &~:=~&	-\frac{6 \left(e_t^2-1\right){}^2}{e_t} \,, \\ 
\hline
\stackrel{(3)}{S}_{31}	&~=~&	0	\,,	\\
\hline
\stackrel{(3)}{S}_{30}	&~=~&	-16 \sqrt{\frac{\pi }{105}} {E}^2 {e_t} \sqrt{1-{e_t}^2} \mu  (1-3 \eta )		\sdelta_{[3 0]} \funcSinAnq{0}{3}\,,	\\
\sdelta_{[3 0]} &~:=~& 1\,, \\
\hline
\stackrel{(4)}{I}_{44}	&~=~&
{
e^{-i\,4\,\peri \Mean}\,
e^{-i 4\phi_0}
\frac{4}{9} \sqrt{\frac{2 \pi }{7}} E^2 \mu  (1-3 \eta ) \times
\left\{
	\sum_{k=0}^{7}		\zeta_{[k4]}	\funcAnq{k}{4}
+	\sum_{k=1}^{7}		\szeta_{[k4]}	\funcSinAnq{k}{4}
\right\}
}\,,
\\
	\zeta_{[0 4]}	&~:=~&	\frac{6 \left(e_t^4-8 e_t^2+8\right)}{e_t^4}	\,, \\ 
	\zeta_{[1 4]}	&~:=~&	-\frac{6 \left(e_t^4-8 e_t^2+8\right)}{e_t^4}	\,, \\ 
	\zeta_{[2 4]}	&~:=~&	-\frac{29 e_t^4-112 e_t^2+88}{e_t^4}	\,, \\ 
	\zeta_{[3 4]}	&~:=~&	\frac{3 e_t^2 \left(e_t^4+8\right)-8}{e_t^4}-19	\,, \\ 
	\zeta_{[4 4]}	&~:=~&	-\frac{32 \left(e_t^2-6\right) \left(e_t^2-1\right){}^2}{e_t^4}	\,, \\ 
	\zeta_{[5 4]}	&~:=~&	\frac{16 \left(e_t^2-1\right){}^3 \left(3 e_t^2-4\right)}{e_t^4} 	\,, \\ 
	\zeta_{[6 4]}	&~:=~&	-\frac{280 \left(e_t^2-1\right){}^4}{e_t^4}	\,, \\ 
	\zeta_{[7 4]}	&~:=~&	-\frac{120 \left(e_t^2-1\right){}^5}{e_t^4}	\,, \\ 
	\szeta_{[1 4]}	&~:=~&	-\frac{24 i \sqrt{1-e_t^2} \left(e_t^2-2\right)}{e_t^3}		\,, \\ 
	\szeta_{[2 4]}	&~:=~&	0	\,, \\ 
	\szeta_{[3 4]}	&~:=~&	-\frac{2 i \sqrt{1-e_t^2} \left(9 e_t^4-46 e_t^2+44\right)}{e_t^3} 	\,, \\ 
	\szeta_{[4 4]}	&~:=~&	-\frac{8 i \sqrt{1-e_t^2} \left(5 e_t^4-17 e_t^2+12\right)}{e_t^3}	\,, \\ 
	\szeta_{[5 4]}	&~:=~&	-\frac{12 i \left(1-e_t^2\right){}^{5/2} \left(e_t^2-8\right)}{e_t^3}	\,, \\ 
	\szeta_{[6 4]}	&~:=~&	\frac{160 i \left(1-e_t^2\right){}^{7/2}}{e_t^3}	\,, \\ 
	\szeta_{[7 4]}	&~:=~&	-\frac{120 i \left(1-e_t^2\right){}^{9/2}}{e_t^3}	\,, \\ 
\hline
\stackrel{(4)}{I}_{43}	&~=~&	0 \,,		\\
\hline
\stackrel{(4)}{I}_{42}	&~=~&	
{
e^{-i\,2\,\peri \Mean} \,
e^{-i 2\phi_0}
\frac{8}{63} \sqrt{2 \pi } E^2 \mu  (1-3 \eta ) \times
\left\{
	\sum_{k=0}^{5}		\zeta_{[k2]}	\funcAnq{k}{2}
+	\sum_{k=1}^{5}		\szeta_{[k2]}	\funcSinAnq{k}{2}
\right\}
}\,,
\\
	\zeta_{[0 2]}	&~:=~&	6-\frac{12}{e_t^2}	\,, \\ 
	\zeta_{[1 2]}	&~:=~&	\frac{12}{e_t^2}-6	\,, \\ 
	\zeta_{[2 2]}	&~:=~&	\frac{16}{e_t^2}-11	\,, \\ 
	\zeta_{[3 2]}	&~:=~&	-\frac{\left(e_t-1\right) \left(e_t+1\right) \left(3  e_t^2-8\right)}{e_t^2}	\,, \\ 
	\zeta_{[4 2]}	&~:=~&	-\frac{20 \left(e_t-1\right){}^2 \left(e_t+1\right){}^2}{e_t^2}	\,, \\ 
	\zeta_{[5 2]}	&~=~&	-\frac{12 \left(e_t-1\right){}^3 \left(e_t+1\right){}^3}{e_t^2} \,, \\
	\szeta_{[1 2]}	&~:=~&	-\frac{12 i \sqrt{1-e_t^2}}{e_t}	\,, \\ 
	\szeta_{[2 2]}	&~:=~&	0	\,, \\ 
	\szeta_{[3 2]}	&~:=~&	-\frac{i \left(3 e_t-4\right) \left(3 e_t+4\right)   \sqrt{1-e_t^2}}{e_t}	\,, \\ 
	\szeta_{[4 2]}	&~:=~&	\frac{8 i \left(1-e_t^2\right){}^{3/2}}{e_t}	\,, \\ 
	\szeta_{[5 2]}	&~:=~&	\frac{8 i \left(1-e_t^2\right){}^{3/2}}{e_t}	\,, \\
\hline
\stackrel{(4)}{I}_{41}	&~=~&	0	\,,	\\
\hline
\stackrel{(4)}{I}_{40}	&~=~&
{
\frac{8}{21} \sqrt{\frac{\pi }{5}} E^2 \mu  (1-3 \eta )\,
\left\{
	\sum_{k=0}^{3}		\zeta_{[k2]}	\funcAnq{k}{2}
\right\}
}
\,,
\\
	\zeta_{[0 0]}	&~:=~&	6			\,,	\\
	\zeta_{[1 0]}	&~:=~&	-6			\,,	\\
	\zeta_{[2 0]}	&~:=~&	-5			\,,	\\
	\zeta_{[3 0]}	&~:=~&	5-5 e_t^2	\,,	\\
\hline \nonumber
\end{eqnarray}
{
It seems that there are divergences for $e_t \rightarrow 0$, but the sum of
the divergent terms is finite.
}
Each decomposition term will contribute to $\sin j\Mean$ and to $\cos j\Mean$,
allowing for $j=[0,\infty)$,
with its own values [$n \magnetic$] in boxy brackets for the multipole moment
having the same special ``magnetic'' number $\magnetic$ and
$n$ labeling contributions with $A(u)^{-n}$.
Using this
and, not less importantly, Eqs. (\ref{Eq::Separate_Ainvnq_M}) and
(\ref{Eq::Separate_SinAinvnq_M}) to separate ${\rm exp}(-i\magnetic\peri\Mean)$
from the products of $A(u)^{-n}$ and $\sin u \, A(u)^{-n}$
with ${\rm exp}(-i\magnetic v_{e_t})$, the moments have the 
Fourier decomposition
\begin{eqnarray}
 \stackrel{(2)}{I}_{22} ~=~&
{
  e^{-2i\phi_0}
  e^{-2i\peri\Mean}
}&
{
  8 \sqrt{\frac{2 \pi }{5}} \,  E \, \mu  \, e_t^{-2} \,
}
\\
&&
{
\Biggl(
	\sum_{j=1}^{\infty}	\sin {j\Mean}	\left\{	\sum_{k=0}^{5} \alpha_{[k2]}  \funcAKEs{k,2}_j +	\sum_{k=1}^{5} \salpha_{[k2]}  \funcSinAKEs{k,2}_j
						\right\}
 } \,, \nonumber 
\label{Eq::I222_Fou_Cons}
\\
 &&
{
+	\sum_{j=0}^{\infty}	\cos {j\Mean}	\left\{	\sum_{k=0}^{5} \alpha_{[k2]}  \funcAKEc{k,2}_j +	\sum_{k=1}^{5} \salpha_{[k2]}  \funcSinAKEc{k,2}_j
						\right\}
\Biggr)
}
\,, \nonumber 
\\
\hline
\stackrel{(2)}{I}_{21}~=~&0\,,
&
\\
\hline
 \stackrel{(2)}{I}_{20} ~=~&
{
}
{
\displaystyle
-16 \sqrt{\frac{\pi }{15}} E \mu
}
&
{
\Biggl(
	\sum_{j=1}^{\infty}	\sin {j\Mean}	 \left\{	\sum_{k=0}^{3} \alpha_{[k2]}  \funcAKEs{k,2}_j 
						\right\}
+
	\sum_{j=0}^{\infty}	\cos {j\Mean}	 \left\{	\sum_{k=0}^{5} \alpha_{[k2]}  \funcAKEc{k,2}_j 
						\right\}
\Biggr)
 }
 \,,  
\\
\hline
\stackrel{(2)}{S}_{22}~=~&0\,,
&
\\
\hline
\stackrel{(2)}{S}_{21}~ =~&
{
  e^{-i\phi_0}
  e^{-i \, \peri\,\Mean}
}&
{
\frac{32}{3} \sqrt{\frac{\pi }{5}} \sqrt{1-e_t^2} \eta  {\mEn}^{3/2} 
({m_1}-{m_2}) \times
}
\\
&&
{
\Biggl(
	\sum_{j=1}^{\infty}	\sin{j\Mean} \left\{	\sum_{k=2}^{3} \beta_{[k1]}  \funcAKEs{k,1}_j +	
																					 \sbeta_{[31]}  \funcSinAKEs{1,1}_j
						\right\}
 }
\nonumber
\\
 &&
{
+	\sum_{j=0}^{\infty}	\cos{j\Mean} \left\{	\sum_{k=2}^{3} \beta_{[k1]}  \funcAKEc{k,1}_j +	
																					\sbeta_{[31]}  \funcSinAKEc{k,1}_j
						\right\}
\Biggr)
}
 \,, \nonumber 
\\
\hline
\stackrel{(2)}{S}_{20}~=~&0\,,
&
\\
\hline
\stackrel{(3)}{I}_{33} ~=~&
{
  e^{-3i\phi_0}
  e^{-3i \, \peri\,\Mean}
}&
{
8 \sqrt{\frac{2 \pi }{21}} (-{E})^{3/2} \left(m_1-m_2\right) \eta \times
}
\\
&& 
\Biggl(
\FourierSin{3}{0}{5}{1}{5}{\gamma}{\sgamma} 
\nonumber \\ &&
+  \FourierCos{3}{0}{5}{1}{5}{\gamma}{\sgamma}
\Biggr)
\,, \nonumber 
\\
\hline
\stackrel{(3)}{I}_{32} ~=~&0\,,
&
\\
\hline
\stackrel{(3)}{I}_{31} ~=~&
{
  e^{-i\phi_0}
  e^{-i \, \peri\,\Mean}
}&
{
8 \sqrt{\frac{2 \pi }{35}} \mEn^{3/2} \left(m_1-m_2\right) \eta \times
}
\\
&&
\Biggl(
 \FourierSin{1}{0}{3}{1}{3}{\gamma}{\sgamma} 
\nonumber \\ &&
+  \FourierCos{1}{0}{3}{1}{3}{\gamma}{\sgamma}
\Biggr)
\,, \nonumber 
\\
\hline
\stackrel{(3)}{I}_{30} ~=~&0\,,
&
\\
\hline
\stackrel{(3)}{S}_{33}~=~&0\,,
&
\\
\hline
 \stackrel{(3)}{S}_{32}~=~&
{
  e^{-2i\phi_0}
  e^{-2i \, \peri\,\Mean}
}&
{
\frac{8}{3} \sqrt{\frac{2 \pi }{7}} E^2 \sqrt{1-e_t^2} \mu  (1-3 \eta ) \times
}
\\
&&
\Biggl(
 \FourierSin{1}{2}{5}{3}{5}{\delta}{\sdelta} 
\nonumber \\ &&
+  \FourierCos{1}{2}{5}{3}{5}{\delta}{\sdelta}
\Biggr)
\,, \nonumber 
\\
\hline
\stackrel{(3)}{S}_{31}~=~&0\,,
&
\\
\hline
\stackrel{(3)}{S}_{30}~=~&&
-16 \sqrt{\frac{\pi }{105}} {E}^2 {e_t} \sqrt{1-{e_t}^2} \mu  (1-3 \eta )	\times	
\nonumber \\&&
\Biggl(
\displaystyle	{\sum_{j=0}^{\infty}} \sin j\Mean
\Biggl\{
	\left( \sdelta_{[3 0]}	\funcSinAKEs{3,0}_j 	\right)
\Biggr\}
+
\displaystyle	\sum_{{j=0}}^{\infty} \cos j\Mean
\Biggl\{
	\left( \sdelta_{[3 0]}	\funcSinAKEc{3,0}_j 	\right)
\Biggr\}
\Biggr)
\,,
\\
\hline
 \stackrel{(4)}{I}_{44}~=~&
{
  e^{-4i\phi_0}
  e^{-4i \, \peri\,\Mean}
}&
{
\frac{4}{9} \sqrt{\frac{2 \pi }{7}} E^2 \mu  (1-3 \eta ) \times
}
\\
&&
\Biggl(
 \FourierSin{4}{0}{7}{1}{7}{\zeta}{\szeta} 
\nonumber \\ &&
+  \FourierCos{4}{0}{7}{1}{7}{\zeta}{\szeta}
\Biggr)
\,, \nonumber 
\\
\hline
\stackrel{(4)}{I}_{43}~=~&0\,,
&
\\
\hline
 \stackrel{(4)}{I}_{42}~=~&
{
  e^{-2i\phi_0}
  e^{-2i \, \peri\,\Mean}
}&
{
\frac{8}{63} \sqrt{2 \pi } E^2 \mu  (1-3 \eta ) \times
}
\\
&&
\Biggl(
 \FourierSin{2}{0}{5}{1}{5}{\zeta}{\szeta} 
\nonumber \\ &&
+  \FourierCos{2}{0}{5}{1}{5}{\zeta}{\szeta}
\Biggr)
\,, \nonumber 
\\
\hline
\stackrel{(4)}{I}_{40}~=~&0\,,
&
\\
\hline
 \stackrel{(4)}{I}_{40}~=~
{
}&
{
\frac{8}{21} \sqrt{\frac{\pi }{5}} \En^2 \mu  (1-3 \eta ) 
}
&
{
\Biggl(
\displaystyle \sum_{j=1}^{\infty} \sin j\Mean
\Biggl\{
\sum_{k=0}^{3} \zeta_{k0} \funcAKEs{k0}_j
\Biggr\}
+\displaystyle \sum_{j=1}^{\infty} \cos j\Mean
\Biggl\{
\sum_{k=0}^{3} \zeta_{k0} \funcAKEc{k0}_j
\Biggr\}
\Biggr)
}
\,.
\label{Eq::I404_Fou_Cons}
\end{eqnarray}
The Fourier transformation of the above gravitational waveforms is easily done.
We take the Fourier transformation of the $\sin j \Mean$ and $\cos j\Mean$ terms,
\begin{eqnarray}
\frac{1}{\sqrt{2 \pi}}
 \int_{-\infty}^{+\infty} 
e^{-i \magnetic \peri \Mean} \sin {j\Mean} \, e^{i \omega t} {\rm d}t&=&
i \sqrt{\frac{\pi }{2}} 
	\delta (j n+\peri \magnetic n-\omega)
-i \sqrt{\frac{\pi }{2}} 
	\delta (j n - \peri \magnetic n+\omega)\,, \\
\frac{1}{\sqrt{2 \pi}}
 \int_{-\infty}^{+\infty} 
 e^{-i \magnetic \peri \Mean} \cos {j\Mean} \, e^{i \omega t} {\rm d}t&=&
\sqrt{\frac{\pi }{2}} \delta (j n+\peri \magnetic n-\omega)
+\sqrt{\frac{\pi }{2}} \delta (j n-\peri \magnetic n+\omega)\,,
\end{eqnarray}
and apply this to the sums over $j$.
 How many terms of the infinite sum combinations are necessary
 to determine the multipoles up to a certain arbitrary order in
 the eccentricity of the binary system is discussed in detail in
 Appendix \ref{Sec::Accuracy}.
Note that in Eqs. (\ref{Eq::I222_Fou_Cons}) to (\ref{Eq::I404_Fou_Cons}), the expansion coefficients
are not longer dependent on $\Mean$. They are functions of
$e_t$, $n$ and the ``magnetic'' number $\magnetic$ and
include multiple sums as we keep in mind the previous sections.
In 1pN-accurate orbital dynamics, they are constants. Due to
RR, they will become slowly varying functions of time. All the
terms, where this condition is satisfied and which will appear
together with an explicit phase factor, can be treated with
the help of the SPA. All other ones which do not have a phase
factor have to be treated individually.

\section{The effect of radiation reaction and the stationary phase approximation of the GW field}
\label{Sec::SPA}
\subsection{Radiative dynamics}
Having the conservative evolution under control,
we can apply the Fourier decomposition to the radiative
evolution. This is done by separating the full orbital
evolution into two time scales: the {\em orbital} and
the {\em reactive} timescale.  The latter is -- for our
calculation -- assumed  to be much larger than the
orbital scale. This is
equivalent to the assumption that the rate of change of
the orbital frequency is small measured over one orbit.
The time dependence of the orbital frequency is closely
related to the loss of energy and orbital angular momentum
due to the GW emission.
Peters and Mathews \cite{Peters:Mathews:1963, Peters:1964} proposed a
relatively simple model of the binary inspiral in the
approximation of slow motion and weak gravitational
interaction for arbitrary eccentricities $0~{\le}~e<1$,
which has been standard for some time. Blanchet, Damour,
Iyer, and Thorne made endeavors to obtain higher-order
corrections to the quadrupole formula
%
\cite{Thorne:1980, Blanchet:1987, Blanchet:Damour:1986, Damour:Iyer:1991}.
The far-zone fluxes of energy and angular momentum
to 2pN order can be found in \cite{Blanchet:Damour:Iyer:1995}.
These corrections have shown up to be unrenounceable
for the data analysis community.
For the time being, we will restrict ourselves to the
Peters-Mathews model for an exemplary calculation. Higher-order
corrections can be included in a forthcoming publication.

In the conservative case, $\MeMo$ and $e_t$ were constants
of motion, but when the orbit shrinks due to RR, both elements
will follow coupled EOM connected to the loss of orbital
energy and angular momentum
\cite{Damour:Gopakumar:Iyer:2004, Konigsdorffer:Gopakumar:2006},
\begin{eqnarray}
	\dot{n}		&=&	{\cal N}(n,e_t)		\,, \label{Eq::EOM_n}\\
	\dot{e}_t		&=&	{\cal E}(n,e_t)		\,, \label{Eq::EOM_e}
\end{eqnarray}
symbolically. Note that above Eqs. hold after averaging over one orbit.
Thus, $\MeMo$ will not simply be $\Mean = \MeMo (t-t_0)$,
but will satisfy
\begin{eqnarray}
 \Mean(t)		&=& \int_{t_0}^{t} \MeMo(t') {\rm d}t'	\,, \\
\dot \Mean(t)	&=& \MeMo(t) \,.
\end{eqnarray}
The Fourier-domain wave forms will not longer follow
Eqs. (\ref{Eq::I222_Fou_Cons}) -- (\ref{Eq::I404_Fou_Cons})
and will require some more considerations.
We will apply the SPA
at the first order to
approximate the frequency-domain wave form first
to each single frequency separately and will later sum up
all the terms of the discrete decomposition as a virtue of
the linearity of the Fourier integral with respect to the
integrand.
\subsection{SPA of the GW signal}
Suppose an integral of the form 
\begin{equation}
\label{Eq::Fou_transform}
 \tilde{ h}(f) = \int_{-\infty}^{+\infty}
	{\cal A}(t) \, e^{i\left(  2\,\pi f t - \phi(t)  \right)} {\rm d}t
\,.
\end{equation}
Then, having found a stationary point $t^*$, where the
phase defined as $\Phi := 2\,\pi f - \phi(t)t$ has zero
{ascent}, $\dot \Phi(t_0) =0 $, with the assumption of
the right behavior at the boundary,  the integral takes
the form

\begin{eqnarray}
 \tilde{h}(f) \approx \sqrt{2 \pi}\frac{A(t^*)}{\sqrt{\ddot \phi(t^*)}} e^{-i \left(\phi(t^*)-2 \pi f t^* + \pi/4\right)}\,.
\end{eqnarray}
In the case of our GW, the integral turns out to be

\begin{eqnarray}
 \stackrel{(n)}{\tilde{h}(f)}_{nm}	&=&	
\frac{1}{\sqrt{2\pi}}
e^{-\magnetic i\phi_0} \int_{-\infty}^{+\infty} 
\sum_{j=0}^{\infty}
 \left(			
S_j \, \sin{j\Mean} + C_j \, \cos{j\Mean}
\right)\,e^{-i \magnetic \peri \Mean}
\, e^{i\,2\pi f t}\,
{\rm d}t
\,,
\\
\stackrel{(n)}{\tilde{h}(f)}_{C, nm}
&=&	\frac{1}{\sqrt{2\pi}}
		e^{-\magnetic i\phi_0}
	{\int_{-\infty}^{+\infty}} \frac{C_j}{2} \left(	
	e^{	ij	\Mean - i \magnetic \peri\Mean}
+	e^{-	ij	\Mean - i \magnetic \peri\Mean}		\right)\,
e^{i2\pi f t}
{\rm d}t
\nonumber
\\
&=&	\frac{1}{\sqrt{2\pi}}
		e^{-\magnetic i\phi_0}
	{\int_{-\infty}^{+\infty}} \frac{C_j}{2} \left(	
   e^{	i\left( j\Mean + 2\pi f t -  \magnetic\peri\Mean\right)	}
+  e^{	i\left(-j\Mean + 2\pi f t -  \magnetic\peri\Mean\right)	}		\right)\,
{\rm d}t
\label{Eq::Fou_hc}
\,,
\end{eqnarray}
where the subscript ``$C$'' shall denote the contribution of $\cos j\Mean$.
The reader should carefully note that $C$ and $S$ are functions
of the elapsed time, as the orbit decays due to RR.
We assumed that the binary system evolves far away from the
last stable orbit, such that the periastron advance parameter is much smaller
than unity, $\peri \ll 1$.
Thus, $n\peri \ll j$ with $j=1,2,3,...$.
{The case $j=0$ will be discussed in one of the upcoming subsections.}
In the first part of Eq. (\ref{Eq::Fou_hc}), therefore, there exist no points
of stationary phase for $j > 0$ and this term vanishes.
The second term in (\ref{Eq::Fou_hc}), having the exponential argument
$i\left(-j\Mean + 2\pi f t	-  \magnetic\peri\Mean\right)$
will contribute, since the phase $\Phi_{mj}$ defined as
\begin{equation}
\label{Eq::Def_Phi}
 \Phi_{mj} :=	2\pi f t - j\Mean	-  \magnetic\peri\Mean
\end{equation}
has a stationary point at $t^*$ where
\begin{eqnarray}
 \dot \Phi_{mj} &=& 2\pi f - j\MeMo - \magnetic \peri\MeMo =0
\,.
\end{eqnarray}
Here we note that $\dot \peri={\cal O}(\epsilon^7)$ and will be 
consistently neglected.
We want to compute $t^*$ and its contribution in the following
lines.
The question arises how to find out $t^*$ without solving
Eqs. (\ref{Eq::EOM_n}) and (\ref{Eq::EOM_e}) numerically. It is answered in a
simple manner. We search for the solution $n=n(e_t, e_{t0})$,
$e_{t0}$ being the eccentricity at $t=t_0$,
of
\begin{equation}
 \frac{{\rm d}n}{{\rm d} e_t}	=	\frac{\dot{n}}{\dot{e}_t}	\,,
\end{equation}
insert it into Eq. (\ref{Eq::EOM_e}),
\begin{eqnarray}
\dot{e}_t
&=&
{\cal E}(n(e_t),e_t, e_{t0})
\,, \\
\Rightarrow
e_t &=&	e_t (e_{t0}, t-t_0)\,, \label{Eq::et_et0_t}
\end{eqnarray}
and invert Eq. (\ref{Eq::et_et0_t}) to find $(t-t_0)$
as a function of $e_t$ and $e_{t0}$, where for the
stationary point the value $e^*_t$ corresponds
to the solution of the equation
\begin{equation}
\label{Eq::stationary_et}
 0=2\pi f - j\MeMo(e_t^*, e_{t0}) - \magnetic \peri(n(e_t^*, e_{t0}), e_t^*) \, n(e_t^*, e_{t0})
\,.
\end{equation}
We see that $e_t^*$ and thus $t^*$ depends only on the
magnetic number $\magnetic$, the summation index $j$,
the eccentricity and mean motion at $t_0$, $e_{t0}$ and
$n_0$, and of course the frequency $f$.
The $\sin$ term of all those calculations will have the same
stationary point $t^*$,
such that the full Fourier-domain wave form will have the following appearance:

\vspace{0.3cm}
\hspace{0cm}
\fbox{
\begin{minipage}[!hr]{0.95\textwidth}
 \begin{eqnarray}
\label{Eq::htt_SPA}
{  \stackrel{(n)}{\tilde{h}}}(f)_{n, \magnetic>0}
 &=&
\epsilon^4 \frac{G}{R}
\sum_{j=0}^{\infty} \frac{1}{2}
\left(
i{S^*_{m (j>0)} }+ (1+\delta_{0j}) C^*_{mj} \right)
\left[ 
\frac{
1
}{\sqrt{\ddot \phi_{mj}(t^*_{mj})}} e^{i \left(\Phi_{m j} (t^*_{m j}) - \pi/4\right)} \right]
e^{-i\magnetic \phi_0}
\label{eq::htt_SPA_m_ge_0}
\,,
\\
{  \stackrel{(n)}{\tilde{h}}}(f)_{n, \magnetic=0}
 &=&
\epsilon^4 \frac{G}{R}
\sum_{j=1}^{\infty} \frac{1}{2}
\left(
i{S^*_{0 j} } + C^*_{0 j} \right)
\left[ 
\frac{
1
}{\sqrt{\ddot \phi_{0j}(t^*_{0j})}} e^{i \left(\Phi_{0 j} (t^*_{0 j}) - \pi/4\right)} \right]
\,,
\\
\ddot{\phi}_{mj} (t_{mj}^*)
		&:=& \dot \MeMo \left( j + \magnetic \peri \right) |_{t=t^*_{mj}}
\,
\\
S_{mj}^*	&:=& S_{mj}		 |_{t=t^*_{mj}}	\,, \\	
C_{mj}^*	&:=& C_{mj}		 |_{t=t^*_{mj}}	\,.	
 \end{eqnarray}
\end{minipage}
}
\vspace{0.3cm}

\noindent
The addend $\delta_{0j}$ appears only at one special place in Eq. (\ref{eq::htt_SPA_m_ge_0}) because in Eq. (\ref{Eq::Fou_hc})
both terms, the first {\em and} the second, contribute to the $\cos$ terms
for $j=0$
and the $\sin$ terms cancel each other.
For computing the energy $E$ which is a usual prefactor with some
exponent in all the multipoles, use of Eq. (\ref{eq::n_of_E}) has to be made
when the ``stationary'' $e_t$ and $n$ are determined.
This approximation is justified as we simply use the
leading-order SPA and is valid in a regime where the
orbit evolves {\em slowly} towards coalescence. It is, thus,
an approximation around the point ($n_0, e_{t0}$) with the
aforementioned requirements.

These are the ideas so far. Here are the explicit equations
up to 1pN in conservative and only to leading-order in radiative
dynamics, where  $n$ is unscaled (that means it has unit second$^{-1}$),

\begin{eqnarray}
\dot{n}
&= {\cal N}(n,e_t)
&=
\epsilon^5
\frac{n^2 \eta  \left(37 e_t^4+292 e_t^2+96\right) (G {\mtot} n)^{5/3}}{5
   \left(1-e_t^2\right){}^{7/2}}
\,,\\
\dot{e}_t
&= {\cal E}(n,e_t)
&=-\epsilon^5
\frac{n \eta  e_t \left(121 e_t^2+304\right) (G {\mtot} n)^{5/3}}{15 \left(1-e_t^2\right){}^{5/2}}
\,.
\end{eqnarray}
In terms of combined quantities, related formulas
for the simple Peters and Matthews approach have
been published by Pierro and Pinto {\cite{Pierro:Pinto:1996}}
and Appell's 2-variable hypergeometric function
${\rm AppellF}_1$  has come to use. We will keep
our own expressions for $n$ and $e_t$ and will
express unscaled elapsed times as functions of the
latter. Note that by an appropriate scaling the
factor $G\,\mtot$ is re-absorbed in the time
unit.

The solution to ${\rm d}\MeMo/{\rm d}e_t = ({\rm d}\MeMo/{\rm d}t)/({\rm d}e_t/{\rm d}t)$ reads
\begin{equation}
\label{Eq:: n_of_et}
 \MeMo(n_0, e_t, e_{t0})
	=	\MeMo_0\,
\left(\frac{\left({e_t}^2-1\right)}	{\left(e_{t0}^2-1\right)}\right)^{3/2}
\left(\frac{ e_{t0}}				{e_t}	\right)^{18/19}
\,
\left(
\frac{ 121  {e_{t0}}^2+304}{ 121  {e_t}^2+304}
\right)^{1305/2299}
\,,
\end{equation}
with $e_{t0}$ and $n_0$ as the value of $e_t$ and $n$ at the initial
instant of time $t_0$, respectively.
The elapsed time as a function of $e_t$ and $e_{t0}$ reads
\begin{eqnarray}
\label{Eq::time_elapsed}
t-t_0 
&=&
\epsilon^{-5} \,
\frac{95 19^{1181/2299}}
{2 2^{2173/2299} {c_0}^{8/3} \eta  (G {\mtot})^{5/3}} \times
\nonumber \\ && 
 \biggl\{{e_{t0}}^{48/19}
   {\rm AppellF}_1\left(\frac{24}{19};\frac{3}{2},-\frac{1181}{2299};\frac{43}{19}; {e_{t0}}^2,-
   \frac{121 {e_{t0}}^2}{304}\right)
\nonumber \\ && 
~
-{e_t}^{48/19}
   {\rm AppellF}_1\left(\frac{24}{19};\frac{3}{2},-\frac{1181}{2299};\frac{43}{19}; {e_t}^2,
-\frac{121 {e_t}^2}{304}\right)\biggr\}
\,.
\\
c_0 &:=&\frac{e_{t0}^{18/19} \left(121 e_{t0}^2+304\right)^{1305/2299}
   {n_0}}{\left(1-e_{t0}^2\right)^{3/2}}\,.
\end{eqnarray}
A check will show that the right hand side has the dimension of time.

\subsection{Solution to the SPA condition equation, $j\neq0$}

It showed up that it is easier to solve Eq. (\ref{Eq::stationary_et}) for $\MeMo$
instead of $e_t$ and then to express $e_t-e_{t0}$ in terms of $\MeMo-\MeMo_0$.
Expressed fully in terms of $\MeMo$, it reads
\begin{equation}
 0=2\pi f - j\MeMo^* - \magnetic \peri(\MeMo^*, e_t(\MeMo^*)) \MeMo^*
\,.
\end{equation}
It is solved in two steps: first, solve the Newtonian and second: solve the 1pN equation with the help of step 1,
\begin{eqnarray}
 {\rm Step ~I:}	& \hspace{2cm}  \MeMo^*_{\rm N}		&=	 \frac{2 \pi f}{j} \,, \\
 {\rm Step ~II:}	& \hspace{2cm}  \MeMo^*_{\rm 1pN}		&=	 \frac{2 \pi f - \magnetic \peri (n^*_{\rm N}, e_t(n^*_{\rm N}) )\, n^*_{\rm N} }{j} \label{Eq::n_stat_j_neq_0}\,.
\end{eqnarray}
Step II drops out in case $\magnetic=0$.
The Newtonian ``stationary eccentricity'' for the frequency $f$,
i.e. $e_t(n^*_{\rm N})$,
can be found numerically or with the help of
a perturbative solution scheme.
{It is a rather numerical issue to apply
fixpoint-method-like iterative algorithms a la
Danby and Burkhards' method \cite{Danby:Burkhardt:1983}
to solve the Kepler
equation, and a detailed analysis could
{indicate} how many steps
are necessary and reasonable towards the solution.
Anyway, we promised to give analytical results
and this will be done below.}
We are aware that there may exist better
algorithms and refer the reader to the common literature of
approximative solving methods.
The function to be inverted for $e_t -e_{t0}$ is
the following,
\begin{eqnarray}
 \MeMo(n_0, e_t, e_{t0})
	&=&
	\MeMo_0\,
\left(\frac{\left({e_t}^2-1\right)}	{\left(e_{t0}^2-1\right)}\right)^{3/2}
\left(\frac{ e_{t0}}				{e_t}	\right)^{18/19}
\,
\left(
\frac{ 121  {e_{t0}}^2+304}{ 121  {e_t}^2+304}
\right)^{1305/2299}
\\
&=& n_0 
+ 			g^{(1)} (e_t - e_{t0})
+ \tilde\epsilon \, 		g^{(2)} (e_t - e_{t0})^2
+ \tilde\epsilon^2 \,  g^{(3)} (e_t - e_{t0})^3
+ \dots
\,,
\\
g^{(p)} &:=& \frac{1}{p!}\,\frac{\partial^p} {\partial_{e_t}^p}   n(n_0, e_t, e_{t0}) |_{e_{t0}}\,,
\end{eqnarray}
having introduced some smallness parameter $\tilde \epsilon$ which will be set $1$ after the calculation.
The solution algorithm reads (defining $\kappa$ as the difference of $e^*_t$ and $e_{t0}$ and leaving out the ``star'')
\begin{eqnarray}
 \kappa^{[N]}		&:=&	(e_t - e_{t0})^{[N]} 	\,,	\\
 \kappa^{[1]}		&=&	\frac{(n - n_0)}{g^{(1)}}	\,,	 \\
 \kappa^{[2]}		&=&	\frac{1}{g^{(1)}}\left\{ (n - n_0) - \tilde\epsilon \,  g^{(2)} \, ( \kappa^{[1]})^2 \right\}	\,,	 \\
 \kappa^{[N]}		&=&	\frac{1}{g^{(1)}}\left\{ (n - n_0) - 
\sum_{p=2}^{N}
\tilde\epsilon^{p-1} \, 
g^{(p)}\,(\kappa^{[N+1-p]})^p
\right\}	\,,
\end{eqnarray}
with some current solution order $[N]$. For convenience, we will give the first four orders of $e_t-e_{t0}$ in terms of $n-n_0$.
With the definitions
\begin{eqnarray}
 f_1	&:=&	-121 e_{\text{t0}}^4-183 e_{\text{t0}}^2+304 \,,	\\
f_2	&:=&	37 e_{\text{t0}}^4+292 e_{\text{t0}}^2+96 \,,
\end{eqnarray}
we obtain
\begin{eqnarray}
\kappa^{[4]} &=& 
- \left(n-n_0\right) \, \frac{f_1 e_{\text{t0}}}{3 f_2 n_0}
\nonumber \\
&&
- \tilde\epsilon \left(n-n_0\right)^2
\frac{f_1  e_{\text{t0}} \left(370 e_{\text{t0}}^8+34401 e_{\text{t0}}^6-131844
   e_{\text{t0}}^4-26720 e_{\text{t0}}^2-56832\right)}{18 f_2^3 n_0^2}
\nonumber \\
&&
-
\tilde\epsilon^2 \, \left(n-n_0\right)^3
\Biggl\{
 f_1  e_{\text{t0}}
 \biggl(
\nonumber \\
&& \quad \quad \quad \quad \quad
410700  e_{\text{t0}}^{16}+76370220 e_{\text{t0}}^{14}
\nonumber \\
&& \quad \quad \quad \quad \quad
+\left(3257592723-48470 f_2\right) e_{\text{t0}}^{12}-3
   \left(4124011 f_2+9090903688\right) e_{\text{t0}}^{10}
\nonumber \\
&& \quad \quad \quad \quad \quad
+168 \left(658637 f_2+276828486\right)
   e_{\text{t0}}^8+\left(9406764288-76265528 f_2\right) e_{\text{t0}}^6
\nonumber \\
&& \quad \quad \quad \quad \quad
+768 \left(131251
   f_2+61327636\right) e_{\text{t0}}^4+18432 \left(267 f_2+494320\right) e_{\text{t0}}^2
\nonumber \\
&& \quad \quad \quad \quad \quad
-7274496 \left(7 f_2-1332\right)\biggr)
\Biggr\}
\nonumber \\ &&
\times \left(162 f_2^5
   n_0^3\right)^{-1}
\nonumber \\
&&
-
\tilde\epsilon^3 \left(n-n_0\right)^4
\Biggl\{f_1 
e_{\text{t0}} \biggl(
\nonumber \\
&&
+759795000   e_{\text{t0}}^{24}
\nonumber \\
&& \quad \quad \quad \quad \quad
+211927360500 e_{\text{t0}}^{22}
\nonumber \\
&& \quad \quad \quad \quad \quad
-1850 \left(96940 f_2-10211818689\right) e_{\text{t0}}^{20}
\nonumber \\
&& \quad \quad \quad \quad \quad
+\left(459468074902815-62450686800 f_2\right) e_{\text{t0}}^{18}
\nonumber \\
&& \quad \quad \quad \quad \quad
+90 \left(135716  f_2^2-42030994737 f_2-75142160154162\right) e_{\text{t0}}^{16}
\nonumber \\
&& \quad \quad \quad \quad \quad
+8 \left(755339319 f_2^2+6763450991635   f_2+3192333977427390\right) e_{\text{t0}}^{14}
\nonumber \\
&& \quad \quad \quad \quad \quad
-3 \left(31474158721 f_2^2+56138837764240   f_2+8745121825435200\right) e_{\text{t0}}^{12}
\nonumber \\
&& \quad \quad \quad \quad \quad
+24 \left(7251390883 f_2^2+4696313592840   f_2+143898051631680\right) e_{\text{t0}}^{10}
\nonumber \\
&& \quad \quad \quad \quad \quad
-8 \left(35529727041 f_2^2+21737777348800   f_2+5492015496046080\right) e_{\text{t0}}^8
\nonumber \\
&& \quad \quad \quad \quad \quad
+384 \left(94956913 f_2^2-19783158080  f_2-34649027828480\right) e_{\text{t0}}^6
\nonumber \\
&& \quad \quad \quad \quad \quad
+30720 \left(2707169 f_2^2+277826496 f_2-683225807616\right)   e_{\text{t0}}^4
\nonumber \\
&& \quad \quad \quad \quad \quad
+18186240 \left(2319 f_2^2+594368 f_2-213546240\right) e_{\text{t0}}^2
\nonumber \\
&& \quad \quad \quad \quad \quad
-1745879040 \left(35 f_2^2-16576 f_2+1577088\right)
\biggr)
\Biggr\}
\nonumber \\ &&
\times\left({1944 f_2^7 n_0^4}\right)^{-1}
+ {\cal O}(\tilde\epsilon^4 \, (n-n_0)^5)
\,.
\end{eqnarray}

It is up to the reader to truncate this to some required order in $e_{t0}$ or to extend it in orders of $\tilde\epsilon$.
Having found the ``stationary'' $\MeMo_{\rm 1pN}^*$ from Eq. (\ref{Eq::n_stat_j_neq_0}) and from $e_t^*(\MeMo^*_{\rm 1pN})$,
one can obtain the associated $t^*$ by inserting this into Eq. (\ref{Eq::time_elapsed}).
This in turn can be inserted into Eq. (\ref{Eq::Def_Phi}) to get the value of the phase at
the stationary point $t^*$. 
The reader should also note that the solution to Eq. (\ref{Eq::stationary_et}) will introduce new
1pN correction terms to the multipole moments of { Eqs. (\ref{Def::I222}) - (\ref{Def::I202}) }.

\subsection{Solution to the SPA condition equation, $j=0$ and $\magnetic \neq 0$. The pure periastron phase shift}

The stationary phase condition for the pure
periastron-dependent terms (those with $j=0$) 
reads
\begin{eqnarray}
\label{Eq::SPA_j=0}
\dot \Phi
	&=& 2\pi f - \magnetic \peri \MeMo \nonumber	\\
	&=& 2\pi f - \magnetic \MeMo(e_t^*)\, 
		3\frac{
				n(e^*_t)	^{2/3}}{1-(e_t^*)^2}
	= 0 \,, \\
\Rightarrow
g(e_t)
	&:=& 2 \pi f - \magnetic \MeMo(e_t^*)\, 3\frac{n(e^*_t)^{2/3}}{1-(e_t^*)^2} =0 \,,
\end{eqnarray}

with $n(e_t)$ taken from Eq. (\ref{Eq:: n_of_et}).
Here, we proceed presenting all quantities
expressed in terms of $e_t$. To find the
solution to this equation analytically, we
find it convenient to
consider the perturbation algorithm from
{Danby \& Burkhardt} \cite{Danby:Burkhardt:1983} to the fourth
order.
We need to have a nice initial guess for $e_t$, which we take from
the first-order expansion of Eq. (\ref{Eq::SPA_j=0}) in $e_t-e_{t0}$,
\begin{eqnarray}
 e_t^{[0]}
&=&
\frac{\frac{\pi  f}{3 m {n_0}^{5/3}} \left({242 e_{{t0}}^7}+124 e_{{t0}}^5-{974 e_{{t0}}^3}+{608  e_{{t0}}}\right)+178 e_{{t0}}^5-669
   e_{{t0}}^3-784 e_{{t0}}}{3 \left(19 e_{{t0}}^4-284
   e_{{t0}}^2-160\right)}
\,,
\end{eqnarray}
noting that the case $\magnetic=0$ is excluded. This can be inserted into
an iterative solution algorithm, which solves for $\delta$ in the expression
\begin{eqnarray}
 e_t^*					&=& e_t^{[0]} + \delta\,, \\
 g(e_t^{[0]} + \delta)	&=& 0\,.
\end{eqnarray}
This $\delta$ is found with the help of the following procedure, 
\begin{eqnarray}
 \delta_1	&=& - \frac{g}{g'} \,, \\
 \delta_2	&=& - \frac{g}{g' + \frac{1}{2} \delta_1 \, g''} \,, \\
 \delta_3	&=& - \frac{g}{g' + \frac{1}{2} \delta_2 \, g'' + \frac{1}{6} \delta_2^2 \, g''' } \,, \\
 \delta_4	&=& - \frac{g}{g' + \frac{1}{2} \delta_3 \, g'' + \frac{1}{6} \delta_3^2 \, g''' +\frac{1}{24} \delta_3^3 g''''} \,, \\
g^p		&:=& \frac{\partial ^p}{\partial e_t ^p} g(e_t)\,.
\end{eqnarray}
We have $e_t^{[4]} = e_t^{[0]} + \delta_4$ as the fourth-order solution to Eq. (\ref{Eq::SPA_j=0})
with quintic convergence,
and again extract $n(e_t^{[4]})$, $t-t_0$ and so on.
The case $m=0$ will be discussed below.

\subsection{The case $j=0$ and $m=0$. Fourier transformation of a slow-in-time signal}

 In Eq. (\ref{Eq::Fou_transform}), there is no fast oscillating term but only a slow variable
of time to be Fourier transformed,
\begin{equation}
 \tilde{ h}(f) = \int_{-\infty}^{+\infty}
	{\cal A}(t) \, e^{i\left( - 2 \pi f t \right)} {\rm d}t
\,.
\end{equation}
The term ${\cal A}(t)$ depends on time only due to RR.
Those terms are nontrivially dependent on time and have
to be treated individually when they are requested
analytically. 
In principle, one would have to express $\En$, $\MeMo$
and $e_t$ as explicit functions of time. That would
include inversion of Appell functions or perturbation
theory.

However, for the case of inspiralling
compact binaries, they will not be able to significantly contribute
to frequencies  in comparison to those
with fast oscillating exponents as we compare typical
time scales for one orbit and for the inspiral.
We will therefore impose the following relation.
\begin{equation}
 \tilde{h}(f)_{\rm static} \ll \tilde{h}(f)_{\rm stationary}\,,
\end{equation}
where $\tilde{h}_{\rm static}$ means all Fourier integrals
over terms where the mean anomaly -- or equivalently, the time --
does not appear explicitly.
We state that the $(j=0, \magnetic =0)$ Fourier domain terms
almost vanish:
\begin{equation}
 \tilde{h}(f)_{[j=0,\magnetic=0]} \approx 0\,,
\end{equation}
for the frequency domain of interest for
the regarded detector. LISA for example, will
hardly see those terms operating near $n_0 \approx 0.001{\rm Hz}$.
Let us give an exemplary number to support this statement.
{The rate of change of the GW frequency $f$ over one year
will be \cite{Tessmer:Gopakumar:2006}
\begin{equation}
 \Delta f_{\rm RR} \sim \frac{1.6 \times 10^{-9}}{(1-e_t^2)^{7/2}}
 \left( \frac{\mtot}{2.8 M_{\odot}} \right)^{\frac{5}{3}}
 \left(  \frac{\eta}{0.25}\right)
\left( \frac{f_r}{10^{-3} {\rm Hz}}\right)^{\frac{11}{3}}
\left( 1+ \frac{73}{24}e_t^2 + \frac{37}{96}e_t^4\right)
{\rm Hz}\,,
\end{equation}
where $f_r$ is the radial frequency, given by $f_r=n(2\pi)^{-1}$.
Let $n = 10^{-3}$, $m_1=m_2=1.4M_{\odot}$  and $e_t = 0.1$.
Then, $\Delta f_{\rm RR} \sim 2{\times}10^{-12}{\rm Hz}$
and the scaled energy loss is $\Delta E_{\rm RR} \sim 3 \times 10^{-13}$.
}

\subsection{The limit $e_t \rightarrow 0$: the quasi-circular case}
The quasi-circular inspiral has been discussed extensively in the literature,
especially in \cite{Yunes:Arun:Berti:Will:2009} which we have oftenly cited.
For further information, see e.g.
\cite{Arun:Iyer:Sathyaprakash:Sundararajan:2005, Damour:Iyer:Sathyaprakash:2000}.
In the limit $e_t \rightarrow 0$, all elements of our calculation simplify drastically.
The following equations,
\begin{eqnarray}
\Mean				&\rightarrow& 			u\,, \label{Eq::KE_Circ} \\
A(u) 					&\rightarrow& 			1\,, \\
f_{l\magnetic}(u)		&\rightarrow& {\cal F}_{l\magnetic} (E)\,, \\
g_{l\magnetic}(u)		&\rightarrow& {\cal G}_{l\magnetic} (E)\,, \\ 
\phi-\phi_0			&\rightarrow&  (1+\peri)\Mean \,, \\
e^{-i\magnetic \phi} 	&\rightarrow& e^{-i\magnetic \phi_0} e^{-i\magnetic (1+\peri)\Mean } \,, \label{Eq::Phase_Circ}
\end{eqnarray}
show that we have a simple prototype for the SPA for all the multipole moments.
Eq. (\ref{Eq::KE_Circ}) is Kepler's equation for quasi-circular orbits.
The infinite summation series in Eq. (\ref{Eq::htt_SPA}) shrink to one single term,
where the phase term has to be replaced by the one in Eq. (\ref{Eq::Phase_Circ}).
This is because the $\sin u$ terms will always have a factor $e_t$ and
vanish in the case in question.
The value of the phase and the angular velocity, the elapsed time and the resulting
SPA integral can be taken one-to-one from \cite{Yunes:Arun:Berti:Will:2009}.

\subsection{Some concluding remarks for the eccentric inspiral templates}

It is interesting to note how many parameters are included in the
wave form. For the relatively simple case of circular inspiral, the
templates used to have
\begin{itemize}
	\item $t_c$, the time to coalescence,
	\item $\phi_c$, some phase instant,
	\item ${\cal M}_c$, the ``chirp mass'',
	\item $\eta$, the symmetric mass ratio (rather important for higher-
			order corrections to the RR effects), and
	\item $i$, the inclination angle of the orbital plane.
\end{itemize}
Because for eccentric orbits, both $e_t$ and $\MeMo$ will dictate
the contribution to infinitely many frequencies already on the purely
conservative level of EOM, both have to be regarded as parameters
for the template. Thus, we have
\begin{itemize}
	\item $\phi_0$, 
	\item $m$, the total mass of the system,
	\item $\eta$, as before,
	\item $i$ as well,
	\item $n_0$ as the value of $n$ at $t_0$,
	\item $e_{to}$ as the value of $e_t$ at $t_0$.
\end{itemize}
Both $n_0$ and $e_{t0}$ will contribute to the time to coalescence,
see \cite{Pierro:Pinto:1996} for the value of what is called ``lifetime''.
The parameter space has grown by one dimension, but the
good news is that, for data analysis considerations, the ambiguity function
is still maximized in view of $\phi_0$ in a considerably simple way
(see \cite{Martel:Poisson:1999} how to do this).

\section{Conclusions and outlook}
In this article, we provided the far-zone GW form,
including 1pN corrections to the orbital dynamics
as well as to the amplitude.
This was done by applying the 1pN accurate QKP to
the conservative dynamics first and to decompose
each term in Fourier modes and in a second step to
solve the Fourier integrals, modified by the
leading-order effect of radiation reaction, with
the help of the SPA method.
The GW field is given in terms of tensor spherical
harmonics and all terms are given in a purely
analytical form, at least as they are solved to
some required order.
The inclusion of the fully analytically solved
Kepler equation in terms of Bessel functions implies
that single and double infinite summations appear.
It is up to the user to restrict those
summations to finite ones, as far as the required
accuracy demands a minimal $j_{max}$ and eccentricity
expansion, due to the detector and other
requirements in question.

It is interesting to note that this approximation
scheme is easily applicable to the case of spinning
compact binaries with aligned spins and orbital angular
momentum, including the leading-order spin-orbit
interaction \cite{Tessmer:Hartung:Schafer:2010}.
For non-aligned spins, the calculation of the
Fourier domain will be structured more complicatedly,
since the precession of the orbital plane will introduce
another typical frequency in addition to the orbital, the
periastron precession and the RR time scale frequencies.

For a future publication it is {intended}
to include higher-order RR terms to the ``quadrupolar''
contribution and, as it is highly demanded, to include
the 2pN and 3pN point particle Hamiltonians into the
dynamics.

\section{Acknowledgments}

We thank Johannes Hartung for useful discussions.
This work was funded by the Deutsche Forschungsgemeinschaft
(DFG) through SFB/TR7 ``Gravitationswellenastronomie'' and
the DLR through ``LISA -- Germany''.

\appendix
\section{Proof of some summation formulas}
\label{App:Proof}
In this appendix, we like to provide some proves
of formulas we only listed in the previous sections.
Throughout the remaining sections the eccentricity $e_t$ we are using is simply called $e$.
Let us start with the inverse scaled relative separation with some arbitrary
positive integer exponent $n$.
 First, we perform a Taylor series expansion in $e$,
\begin{eqnarray}
 \frac{1}{(1-e \cos u)^n}
	&=& 1+ \sum_{m=1}^{\infty} \frac{(n + m -1)!}{{(n-1)!}} \, \frac{e^m}{m!} \, \cos^m u \nonumber \\
	&=& 1+ \sum_{m=1}^{\infty} \frac{(n + m -1)!}{{(n-1)!}} \, \frac{e^m}{m!} \,	 \left( \frac{1}{2^m} \, \sum_{l=0}^{m}
														\binom{m}{l} \cos (u [m-2 l]) \right)\,,
\end{eqnarray}
and list  the ``factorial'' function of the integer number $n$ as
\begin{equation}
\frac{(n + m - 1)!}{{(n-1)!}} \equiv \prod_{k=1}^{m} (n + k - 1)
\,.
\end{equation}
To optically simplify this equation, we summarize
the terms before $\cos \sim u$ with $\beta^{(n)}_{m,k}$
as follows,
\begin{equation}
 \label{Eq::Def_betanmk}
\beta^{(n)}_{m,k} := \frac{(n+m-1)!}{(n-1)!} \, \frac{1}{m!}\, \frac{e^m}{2^m}
\binom{m}{k}
\,,
\end{equation}
and write the sum with this definition:
\begin{eqnarray}
 \label{Aexpn}
\frac{1}{(1-e \cos u)^n} &=&
1+\sum_{m=1}^{\infty} \sum_{k=0}^{m}
\beta^{(n)}_{m,k} \cos([m-2k]u)
\,.
\end{eqnarray}
To further markably reduce the complexity of this double sum,
it is the task to find out which pairs of $(m,k)$ lead
to the same frequency $ju$ and which $\beta^{(n)}_{m,k}$
have to be added to this frequency contribution:
\begin{eqnarray}
 				|m-2k|&=& j 				\,,\\
\Rightarrow \hspace{1cm}	 m_1 &=& 2k+j 	\,,\\
\Rightarrow \hspace{1cm}	 m_2 &=& 2k-j	\,.
\end{eqnarray}
A small table for the $\cos$ function argument
may help (note that $k\le m$ always holds):
 \begin{table}[h!]
  \begin{tabular}{r|rrrrrr}
m : k 	  & 0 & 1 & 2 & 3 & 4 & 5 \\
\hline
	1 & 1 &-1 &   &   &   &   \\
	2 & 2 & 0 &-2 &   &   &   \\
	3 & 3 & 1 &-1 &-3 &   &   \\
	4 & 4 & 2 & 0 &-2 &-4 &   \\
	5 & 5 & 3 & 1 &-1 &-3 &-5
 \end{tabular}
 \end{table}

\noindent
The zero mode $j=0$ always appears at even numbers
$m$, $m=2i {~\rm with~} i=(1,2,...)$,
and
$m-2k=2i-2k=0$
is satisfied by $k=i$.
Thus, we choose
\begin{equation}
\label{Eq::Def_b0}
b^{(n)}_{0} = \sum_{i=1}^{\infty} \beta^{(n)}_{2i,i}
\,.
\end{equation}
The other frequencies $j>0$, appearing at $m=j+2i$
with $i=(0,1,2,...)$
lead to $k=i$ as well as $k=i+j$,
such that
\begin{equation}
\label{Eq::Def_bj}
 b^{(n)}_j = \sum_{i=0}^{\infty} \beta^{(n)}_{j+2i,i} + \beta^{(n)}_{j+2i, j+i}
\,.
\end{equation}
Summarizing the elements for the first term, we have
\begin{equation}
 \frac{1}{(1-e \cos u)^n} = 1 + b^{(n)}_0 + \sum_{j=1}^{\infty} b^{(n)}_j \cos ju
\end{equation}
with the $b^{(n)}_j$ defined in Eqs. (\ref{Eq::Def_b0}) and (\ref{Eq::Def_bj}).
Having this at hands, it is easy to compute the associated
decomposition of
\begin{eqnarray}
\label{Eq::SinU_Aexpn}
  \frac{\sin u}{(1-e \, \cos u)^n}
	&=& \left( 1 + b^{(n)}_0 +  \sum_{j=1}^{\infty} {b}^{(n)}_j \, \cos ju
		\right) \, \sin u \nonumber \\
	&=&  (1+b^{(n)}_{0}) \sin u + \frac{1}{2}\sum_{j=1}^{\infty} b^{(n)}_j \left( 
					\sin[ (1-j) u] + \sin[(1+j) u]
					\right)
\,.
\end{eqnarray}
This decomposition demands collecting the terms having the
same frequency as well.
Having $m$ times u in the $\sin$
argument,
following $j$ will lead to $\sin(m u)$ and $-\sin(m u)$ for
$\sin(1-j) u$ in the first two lines and for
$\sin(1+j) u$ in the third:
\begin{eqnarray}
 m	&=	(1-j)	\rightarrow	j=1-m	\,, \\
-m	&=	(1-j)	\rightarrow	j=1+m	\,, \\
 m	&=	(1+j)	\rightarrow	j=m-1\,.
\end{eqnarray}
Again, a small table may help:
\begin{table}[!ht]
 \begin{tabular}{c|cccc}
 j & 1 & 2 & 3 & 4 \\
\hline
1+j& 2 & 3 & 4 & 5 \\
1-j& 0 &-1 &-2 &-3
\end{tabular}
\end{table}

\noindent
We clearly see that the first harmonic in this sum is only
realized by $j=2$ for the $\sin (1-j)u$ term, whereas $j=0$
will not contribute. Thus,
\begin{equation}
 \sum_{j=1}^{\infty} b^{(n)}_j \left( \sin[ (1-j) u] + \sin[(1+j) u]	\right) =
 \sum_{m=2}^{\infty} b^{(n)}_{m-1} \sin mu
-\sum_{m=2}^{\infty} b^{(n)}_{m+1} \sin mu
-b^{(n)}_2 \sin u
\,.
\end{equation}
Summarizing these terms, we write the lhs of Eq. (\ref{Eq::SinU_Aexpn}) as
a simple sum.
\begin{eqnarray}
 \frac{\sin u}{(1-e \cos u)^n} &=&
\sum_{j=1}^{\infty}
\SA{n}{j} \sin ju \,, 							\\
\SA{n}{1} &:=& 1 + b^{(n)}_0 - \frac{1}{2} b^{(n)}_{2} \,,		\\
\SA{n}{j>1} &:=& \frac{1}{2} \left( b^{(n)}_{j+1} - b^{(n)}_{j-1} \right)
\,.
\end{eqnarray}

\section{Fourier representation of the $\sin mu$ and $\cos mu$ terms}
In the previous appendix, we have proven some formulas
for terms appearing in the multipole expansion of the
far-zone gravitational field.
Now, we have to express simple series in $\sin mu$ and
$\cos mu$ as trigonometric functions in $ml$.
Using Eq. (\ref{Eq::cosmE}), we have
\begin{eqnarray}
\frac{1}{(1-e \cos u)^n}
&=&	1+b^{(n)}_{0} + \sum_{m=1}^{\infty} b^{(n)}_{m} \cos {mu}										
 =		1+b^{(n)}_{0} + \sum_{m=1}^{\infty} b^{(n)}_{m} \left( \sum_{j=1}^{\infty} \CKE{m}{j} \cos jl \right)	\nonumber \\
&=&	1+b^{(n)}_{0} + \sum_{j=1}^{\infty} \left( \sum_{m=1}^{\infty} \CKE{m}{j} b^{(n)}_{m} \right) \cos jl	\nonumber \\
&=&	\AKE{n}{0} + \sum_{j=1}^{\infty} \AKE{n}{j} \cos {j l} \,, \\
\AKE{n}{0}	&:=&	1+b^{(n)}_{0} \,, \\
\AKE{n}{j>0}	&:=&	\left( \sum_{m=1}^{\infty} \CKE{m}{j} b^{(n)}_{m} \right)
\,.
\end{eqnarray}
Similarly, getting help from Eq. (\ref{Eq::sinmE}), we obtain
\begin{eqnarray}
 \frac{\sin u}{(1-e \cos u)^n}	&=&
 \sum_{m=1}^{\infty} \SA{n}{m} \sin m u = \sum_{m=1}^{\infty} \SA{n}{m} \left( \sum_{j=1}^{\infty} \SKE{m}{j} \sin jl \right)
= \sum_{j=1}^{\infty} \left( \sum_{m=1}^{\infty} \SKE{m}{j} \SA{n}{m} \right) \cos jl		\nonumber	\\
&=& \sum_{j=1}^{\infty} \SAKE{n}{j} \sin jl													\,, \\
 \SAKE{n}{j>0} &:=& \left( \sum_{m=1}^{\infty} \SKE{m}{j} \SA{n}{m} \right)
\,.
\end{eqnarray}

\section{Fourier representation of products of two $\sin$ series}
\label{App::ProdSinSin}

In section \ref{Sec::decomp_elements}, we provided a simple series representation of the term
\begin{eqnarray}
 \left(
\sum_{k=1}^{\infty}
	\SAKE{n}{k} \sin k\Mean
\right)
\left(
\sum_{m=1}^{\infty}
	\vKE{m} \sin m\Mean
\right)
&=&\sum_{k=1}^{\infty} \sum_{m=1}^{\infty}
\frac{1}{2} 
\SAKE{n}{k} \vKE{m}
\left(
\cos[k-m] \Mean - \cos[k+m] \Mean
\right)
\nonumber\\
&=& \sum_{j=0}^{\infty} \ProdSS{j} \cos j\Mean
\,.
\end{eqnarray}
Here we want to explain how the coefficients $\ProdSS{j}$ come up.
First, we take a look at $\cos [k-m]\Mean$.
The zero mode appears only at $k=m$. Thus, the absolute static
part will be given by
\begin{equation}
 \ProdSS{0} := \frac{1}{2}\sum_{m=1}^{\infty} \SAKE{n}{m} \vKE{m} \,.
\end{equation}
The higher harmonics $j>0$ will be realized by $k-m=j$ and $k-m=-j$
and these equations solved for m give
$m=k-j$ (which only nonzero for $k>j$) and $m=k+j$.
The first part, $\cos [k-m]\Mean$, will give
\begin{equation}
 ^{[1]} \ProdSS{j>0} = \frac{1}{2}\sum_{k=1}^{\infty} \left( \underbrace{\SAKE{n}{k} \vKE{k-j}}_{k>j} + \SAKE{n}{k} \vKE{k+j} \right)
\,.
\end{equation}
The second part, $-\cos ([k+m]\Mean)$, gives $-\cos j\Mean$ for $j=k+m$, viz. $m=j-k$,
such that it is equal to
\begin{equation}
  ^{[2]} \ProdSS{j>1} = -\frac{1}{2}\sum_{k=1}^{j-1} \left( \SAKE{n}{k} \vKE{j-k} \right)
\,,
\end{equation}
together giving
\begin{equation}
\ProdSS{j>0} = ~^{[1]}\ProdSS{j>0} + ~^{[2]}\ProdSS{j>1}
=
\frac{1}{2}	\sum_{k=1}^{\infty} \left( \underbrace{\SAKE{n}{k} \vKE{k-j}}_{k>j} + \SAKE{n}{k} \vKE{k+j} \right)
-\frac{1}{2}	\sum_{k=1}^{j-1} \left( \SAKE{n}{k} \vKE{j-k} \right)
\,.
\end{equation}

%
%
%
Additionally, we will provide the tables for part 1 and 2.
\begin{table}[!ht]
\begin{tabular}{ll}
 \begin{tabular}{c|cccc}
 k:m & 1 & 2 & 3 & 4 \\
\hline
 1	& 0 &-1 &-2 &-3 \\
 2	& 1 & 0 &-1 &-2  \\
 3	& 2 & 1 & 0 &-1  \\
 4	& 3 & 2 & 1 & 0  \\
\end{tabular}
&~~~
 \begin{tabular}{c|cccc}
 k:m & 1 & 2 & 3 & 4 \\
\hline
 1	& 2 & 3 & 4 & 5 \\
 2	& 3 & 4 & 5 & 6  \\
 3	& 4 & 5 & 6 & 7  \\
 4	& 5 & 6 & 1 & 8  \\
\end{tabular}
\\
Numbers for part 1
~~~&~~~
Numbers for part2
\end{tabular}
\end{table}

\newpage

\section{Fourier representation of products of a sin and a cos series}
\label{App::ProdSinCos}

What we did in the previous appendix is also necessary for the following term,
\begin{eqnarray}
 \left(
\sum_{k=1}^{\infty} \AKE{n}{k} \cos {k\Mean}
\right)
 \left(
\sum_{k=1}^{\infty} \vKE{k} \sin{k\Mean}
\right)
&=&
\frac{1}{2}
\sum_{k=1}^{\infty}
\sum_{m=1}^{\infty}
\left(
\sin [k+m]\Mean
-\sin [k-m]\Mean
\right)
\nonumber \\
= \sum_{j=1}^{\infty} \ProdCS{j} \sin {j\Mean}
\,.
\end{eqnarray}
Part 1, $\sin [k+m]l$ will contribute to frequencies $j=k+m$ for $k,m > 0$. We
already know the result from part 2 of the last section.
\begin{equation}
^{[1]}\ProdCS{j>1} = \sum_{k=1}^{j-1} \AKE{n}{k} \vKE{j-k}
\end{equation}
Part 2, $-\sin [k-m]l$ will also have positive and negative contributions,
for
$m=k-j$   (implying $k>j$)
$m=k+j$,
giving
\begin{equation}
 ^{[2]}\ProdCS{j>1} = -\frac{1}{2}\sum_{k=1}^{\infty}	\left( \underbrace{\AKE{n}{k} \vKE{k-j}}_{{\rm for~}k>j}-\AKE{n}{k} \vKE{k+j} \right)
\,.
\end{equation}
Summation of part 1 and 2 yields
\begin{eqnarray}
 \ProdCS{j} :=
\frac{1}{2}
\left\{
\underbrace{
  \sum_{k=1}^{j-1}	\left( \AKE{n}{k} \vKE{j-k} \right)
}_{{\rm for~} j>1}
- \sum_{k=1}^{\infty}	\left( \underbrace{\AKE{n}{k} \vKE{k-j}}_{{\rm for~}k>j}-\AKE{n}{k} \vKE{k+j} \right)
\right\}
\,.
\end{eqnarray}

\section{Appell's integral formula in the solution for the elapsed time}

The integral in  Eq. (\ref{Eq::time_elapsed}) in section \ref{Sec::SPA} can be solved with the help of
the following integral representation of the AppellF1 function (see http://dlmf.nist.gov/ or
\cite{Olver:Lozier:Boisvert:Clark:2010} for further information).
\begin{equation}
 \int_{0}^{1} {\rm d}u \, \frac{u^{\alpha-1} (1-u)^{\gamma-\alpha-1}}{(1-u x)^{\beta_1} (1-u y )^{\beta_2}}  =
\frac{\Gamma(\alpha) \Gamma(\gamma -\alpha)}{\Gamma(\gamma)}
{\rm AppellF}_1 \left(
\alpha; \beta_1, \beta_2; \gamma; x , y
 \right)
\,.
\end{equation}

 \section{Accuracy of finite sums}
\label{Sec::Accuracy}
 In Sections \ref{Sec::Ingredients} and  \ref{Sec::decomp_elements}
 we provided decompositions of functions of some ``elementary'' type
 in terms of $u$ which contain infinite summations. Naturally, for numerics it is important
 to know how many terms are needed to reach some desired accuracy.
Considering compact binaries with small eccentricities only, one is allowed to expand the elementary
expressions in powers of $e$ and then to look how many terms are needed for the error to be shifted
to ${\cal O}(e^{\Max+1})$ with some finite $\Max$.

\subsection{Accuracy of basic elements}
We start with the basic definitions. The upper limits of the summation has to give a term of order ${\cal O}(e^\Max)$,
thus the individual limit has to be matched appropriately,
\begin{eqnarray}
\beta^{(n)}_{m,k}	 &=&	\frac{(n+m-1)!}{(n-1)!} \frac{1}{m!} \frac{e^m}{2^m} \binom{m}{k} 	=	{\cal O}(e^m)		\,, \\
 b^{(n)}_0 		 &=&	1+\sum_{i=1}^{\infty} \beta^{(n)}_{2i,i}					=	1+\sum_{i=0}^{\Max/2} \beta^{(n)}_{2i,i}+{\cal O}(e^{\Max+1})	 \,, \\
b^{(n)}_{j>0}	 &=&	\sum_{i=0}^{\infty}\left( \beta^{(n)}_{j+2i,i} + \beta^{(n)}_{j+2i,j+i} \right) 
						= \sum_{i=0}^{(\Max-j)/2}\left( \beta^{(n)}_{j+2i,i} + \beta^{(n)}_{j+2i,j+i} \right) +{\cal O}(e^{\Max+1})\,,		\label{Def::bnj_trunc}		\\
\Rightarrow b^{(n)}_{j>0} &=& {\cal O}(e^j) \,, \\
A(u)^{-n}		&=&	\sum_{j=0}^{\Max} b^{(n)}_{j} \cos ju+{\cal O}(e^{\Max+1}) \,. \label{Eq:Au_cos_ju_finite}
\end{eqnarray}
In the last line we have used that the summation in Eq. (\ref{Def::bnj_trunc}) starts with $i=0$ and leaves no term if $j >\Max$.
The same quantity with $\sin u$ will also be truncated in the $u$ domain, keeping in mind the definitions (\ref{Def::SA_1}) and
(\ref{Def::SAj>1}) and their dependency on the summation index $j$,
\begin{eqnarray}
 \frac{\sin u}{A(u)^{n}}  &=&	\sum_{j=1}^{\infty} \SA{n}{j} \sin ju\,, \\
\SA{n}{1}		&=&	\left( 1+b^{(n)}_{0} \right) - \frac{1}{2} b^{(n)}_{2} = \left(1 + \sum_{i=1}^{\Max/2}\beta^{(n)}_{2i,i}\right)
						-\frac{1}{2}\sum_{i=0}^{(\Max-2)/2}\left( \beta^{(n)}_{2+2i,i} + \beta^{(n)}_{2+2i,2+i} \right)+{\cal O}(e^{\Max+1})	 \,, \\
\SA{n}{j>1}		&=&	\frac{1}{2} \left( 
										\sum_{i=0}^{(\Max-j+1)/2}\left( \beta^{(n)}_{j-1+2i,i} + \beta^{(n)}_{j-1+2i,j-1+i} \right)-
										\sum_{i=0}^{(\Max-j-1)/2}\left( \beta^{(n)}_{j+1+2i,i} + \beta^{(n)}_{j+1+2i,j+1+i} \right)
								\right)
					\nonumber \\ && +{\cal O}(e^{\Max+1})\,, \\
\Rightarrow \SA{n}{j>1}
				&=&	{\cal O}(e^{j-1}) + {\cal O}(e^{j+1}) = {\cal O}(e^{j-1}) \,, \label{Eq::Order_SA}\\
 \frac{\sin u}{A(u)^{n}}  &=& \sum_{j=1}^{\Max+1} \SA{n}{j} \sin ju +{\cal O}(e^{\Max+1}) \,.\label{Eq:SinuAu_cos_ju_finite}
\end{eqnarray}
For the Fourier representation, we remember Eq. (\ref{Eq::Au_KE}) and take the expansion of the Bessel coefficients
\cite{Colwell:1993},
\begin{equation}
 J_{n}(x) = x^n \, \sum_{k=0}^{\infty} \frac{(-1)^k \, x^{2k}}
									{2^{2k+n} k! \, (k+n)!}\,,
\end{equation}
for the determination of the limit for our finite sums,
\begin{eqnarray}
\CKE{m}{j}		&=&	\frac{m}{j} \, \left( J_{j-m}(je) {-} J_{j+m}(je) \right) 
				=    {\cal O}(e^{j-m})     \,,\\
\Rightarrow
\cos mu			&=&	\sum_{j=1}^{\Max+m} \CKE{m}{j} \cos j\Mean -{\cal O}(e^{\Max+1})\,, \\
 \AKE{n}{j}		&=&	\sum_{m=1}^{\Max} \gamma_{j}^{m} b^{(n)}_{m} = {\cal O}(e^{j-m}) \, {\cal O}(e^m)={\cal O}(e^j)	 \label{Eq::order_AKE}	\,,\\
A(u)^{-n}		&=&	(1+b^{(n)}_0) + \sum_{j=1}^{\Max} \AKE{n}{j} \cos j\Mean +{\cal O}(e^{\Max+1}) \,.
\end{eqnarray}
In the third line above we used Eq. (\ref{Eq:Au_cos_ju_finite}) to truncate the number $j$ in $\cos j u$. For the other relevant term with $\sin u$ we consider
\begin{eqnarray}
\SKE{m}{j}			&=&	\frac{m}{j} \left( J_{j-m}(je) {+}  J_{j+m}(je) \right) = {\cal O}(e^{j-m})\,,\\
\Rightarrow
\sin mu				&=&	\sum_{j=1}^{\Max+m} \SAKE{m}{j} \sin j\Mean+{\cal O}(e^{\Max+1}) \,, 											\\
\SAKE{n}{j}			&=&	\sum_{m=1}^{\Max+1}  \SKE{m}{j} \SA{n}{m} =	{\cal O}(e^{j-m}) \, {\cal O}(e^{m-1})	= {\cal O}(e^{j-1}) \,, 	\label{Eq::order_SAKE}	\\
\frac{\sin u}{ A(u)^{n}}		&=&	\sum_{j=1}^{\Max +1 } \SAKE{n}{j} \sin{j\Mean}+{\cal O}(e^{\Max+1}) \,.
\end{eqnarray}
In the third line again, we used (\ref{Eq::Order_SA}) for the index $j$ in $\sin j u$ and in the last Eq. (\ref{Eq:SinuAu_cos_ju_finite}).
We will also consider the expansion coefficients of $v$  from Eq. (\ref{Def::G_in_J}). With the regular solution to Eq.
(\ref{Def::alpha})  at $e=0$,
\begin{equation}
\alpha	= \frac{1-\sqrt{1-e^2}}{e} = \order{e}{1}\,, 
\end{equation}
their order is calculated to be
\begin{eqnarray}
 G_m(e)	&=& \frac{2}{m} J_m (m e) + \sum_{s=1}^{\infty} \alpha^s \left[ J_{m-s}(me) - J_{m+s}(me)\right] \nonumber \\
		&=& \order{e}{m} + \sum_{s=1}^{\infty} \order{e}{s} \left[ \order{e}{m-s} - \order{e}{m+s} \right] \nonumber \\
		&=& \order{e}{m}\,. 
\end{eqnarray}

\subsection{Accuracy of finite sum products}
\noindent
The product of two sin series can be truncated taking into account the
individual series' terms. In the following we will use the fact that those coefficients could
already be truncated to some finite order $\Max$ in the previous subsection,
see Eqs. (\ref{Eq::order_AKE}) and (\ref{Eq::order_SAKE}), and that they are
at least of some order of $e$ themselves,
\begin{eqnarray}
 \left(
\sum_{k=1}^{\infty}
	\SAKE{n}{k} \sin k \Mean
\right)
\left(
\sum_{m=1}^{\infty}
	\vKE{m} \sin m \Mean
\right)
&=& \sum_{j=0}^{\infty} \ProdSS{j} \cos j \Mean
\,.
\end{eqnarray}
For clarity, we again write down the definitions and make use of the elaborated orders,
where some abuse of notation is made,
\begin{eqnarray}
\ProdSS{0} &:=& \frac{1}{2}
\sum_{k=1}^{\infty} \SAKE{n}{k} \vKE{k} \,, \\
\AKE{n}{k}	&=&\order{e}{k}\,,\\
\SAKE{n}{k}	&=&\order{e}{k-1}\,, \\
\vKE{k}		&=&\order{e}{k}\,, \\
\Rightarrow
\ProdSS{0}	&=& \frac{1}{2} \sum_{k=1}^{\infty} \order{e}{k-1} \order{e}{k} = \frac{1}{2} \sum_{k=1}^{\infty} \order{e}{2k-1}  
\stackrel{
=
}
{
\text{\tiny (2k-1 = \Max)}
}
\frac{1}{2}\sum_{k=1}^{\frac{\Max+1}{2}} \SAKE{n}{k} \vKE{k} +\order{e}{\Max+1} = \order{e}{1} \,, 							\label{Eq::order_ProdSS0}\\
\ProdSS{j} &:=& \frac{1}{2} \left\{
\sum_{k=1}^{\infty}
\left(
 \SAKE{n}{k} \vKE{k+j} +  \underbrace{\SAKE{n}{k} \vKE{k-j}}_{{\rm for ~} k>j}
\right)
-
\underbrace{
\sum_{k=1}^{j-1} \SAKE{n}{k} \vKE{j-k}
}_{{\rm for ~} j>1}
\right\}
\nonumber \\
&=&
\frac{1}{2}\left\{
\sum_{k=1}^{\infty} \order{e}{k-1} \order{e}{k+j}
+\sum_{k=j+1}^{\infty} \order{e}{k-1} \order{e}{k-j}
+\sum_{k=1}^{j-1} \order{e}{k-1} \order{e}{j-k}
\right\}
\nonumber\\
&=&
\frac{1}{2}\left\{
\sum_{k=1}^{\infty} \order{e}{2k+j-1}
+\sum_{k=j+1}^{\infty} \order{e}{2k-j-1}
-\sum_{k=1}^{j-1} \order{e}{j-1}
\right\}
\nonumber\\
&=&
\frac{1}{2}\left\{
\sum_{k=1}^{\frac{\Max-j+1}{2}}  \SAKE{n}{k} \vKE{k+j}
+\sum_{k=j+1}^{\frac{\Max +j+1}{2}} \SAKE{n}{k} \vKE{k-j}
-\sum_{k=1}^{j-1}\SAKE{n}{k} \vKE{j-k}
\right\}
+\order{e}{\Max+1}
\nonumber \\
&=&
\order{e}{j-1}
\,. \label{Eq::order_ProdSSj}
\end{eqnarray}
The individual final indices are evaluated when the maximal exponent
of $e$ reaches $\Max$. The last line is evaluated when one takes the
smallest index $k$.
%
\noindent
The other product of $\cos$ and $\sin$ series is the following,
\begin{eqnarray}
 \left(
\sum_{k=1}^{\infty} \AKE{n}{k} \cos {k \Mean}
\right)
 \left(
\sum_{m=1}^{\infty} \vKE{m} \sin{m \Mean}
\right)
&=&
 \sum_{j=1}^{\infty} \ProdCS{j} \sin {j \Mean}
\,.
\end{eqnarray}
The definitions are worked through immediately, again with
some minor abuse of notation,
\begin{eqnarray}
 \ProdCS{j}
&:=&
\frac{1}{2}\left\{
\underbrace{
			\sum_{k=1}^{j-1} \AKE{n}{k} \vKE{j-k}
			}_{{\rm for~} j>1}
- \sum_{k=j+1}^{\infty}	 \AKE{n}{k} \vKE{k-j}
+\sum_{k=1}^{\infty}  \AKE{n}{k} \vKE{k+j}
\right\}
\nonumber \\
&=&
\frac{1}{2}
\left\{
\underbrace{
  \sum_{k=1}^{j-1}	\order{e}{k} \order{e}{j-k}
}_{{\rm for~} j>1}
- \sum_{k=j+1}^{\infty}	 \order{e}{k} \order{e}{k-j}
+\sum_{k=1}^{\infty}  \order{e}{k} \order{e}{k+j}
\right\}
\nonumber \\
&=&
\frac{1}{2}
\left\{
\underbrace{
  \sum_{k=1}^{j-1}	 \order{e}{j}
}_{{\rm for~} j>1}
- \sum_{k=j+1}^{\infty}	 \order{e}{2k-j}
+\sum_{k=1}^{\infty} \order{e}{2k+j}
\right\}
\nonumber \\
&=&
\frac{1}{2}\left\{
\underbrace{
			\sum_{k=1}^{j-1}  \AKE{n}{k} \vKE{j-k}
			}_{{\rm for~} j>1}
- \sum_{k=j+1}^{\frac{\Max+j}{2}}	 \AKE{n}{k} \vKE{k-j}
+\sum_{k=1}^{\frac{\Max-j}{2}}  \AKE{n}{k} \vKE{k+j}
\right\}
\nonumber \\
&=&
\order{e}{j}
\,.
\label{Eq::order_ProdSCj}
\end{eqnarray}
Because of Eqs. (\ref{Eq::order_ProdSSj}) and (\ref{Eq::order_ProdSCj}), one can truncate
the summations over the index $j$ for
\begin{enumerate}
\item each $\AKE{n}{j}$ term at $j_\Max=\Max$,
\item each $\SAKE{n}{j}$ term at $j_\Max = \Max+1$,
\item each $\ProdSS{j}$ term at $j_\Max = \Max+1$, and
\item each $\ProdCS{j}$ term at $j_\Max = \Max$.
\end{enumerate}
in Eqs. (\ref{Eq::I222_Fou_Cons}) to (\ref{Eq::I404_Fou_Cons}).

\end{document}